\documentclass[pra,showpacs,twocolumn,superscriptaddress,floatfix,aps]{revtex4-2}
\usepackage{amsmath,amssymb,bm}
\usepackage{graphicx}
\usepackage{physics}
\usepackage[T1]{fontenc}
\usepackage{comment}
\usepackage{float}
\usepackage{mathtools}
\usepackage[normalem]{ulem}
\usepackage{hyperref}
\hypersetup{colorlinks,%,%
 linkcolor=blue,%
 citecolor=blue,%
 urlcolor=blue}
\usepackage{color}
\usepackage{tikz}
\usetikzlibrary{shapes}
\usepackage{relsize}

\begin{document}
\title{Signature of localization-delocalization in collisional inhomogeneous spin-orbit coupled condensates}			

\author{Swarup K. Sarkar}
%\email{skanti@iitg.ac.in}
\affiliation{Department of Physics, Indian Institute of Technology Guwahati, Guwahati 781039, Assam, India}

\author{Rajamanickam Ravisankar}
\affiliation{Department of Physics, Zhejiang Normal University, Jinhua 321004, PR China}
\affiliation{Zhejiang Institute of Photoelectronics $\&$ Zhejiang Institute for Advanced Light Source, Zhejiang Normal University, Jinhua, Zhejiang 321004, China}

\author{Tapan Mishra}
%\email{mishratapan@gmail.com}
\affiliation{School of Physical Sciences, National Institute of Science Education and Research, Jatni 752050, India}
\affiliation{Homi Bhabha National Institute, Training School Complex, Anushaktinagar, Mumbai 400094, India}

\author{Paulsamy Muruganandam}
%\email{anand@bdu.ac.in}
\affiliation{Department of Physics, Bharathidasan University, Tiruchirappalli 620024, Tamilnadu, India}

\author{Pankaj Mishra}
%\email{pankaj.mishra@iitg.ac.in}
\affiliation{Department of Physics, Indian Institute of Technology Guwahati, Guwahati 781039, Assam, India}
\date{\today}

\begin{abstract}
We study the localization transition in spin-orbit (SO) coupled binary Bose-Einstein condensates (BECs) with collisional inhomogeneous interaction trapped in a one-dimensional quasiperiodic potential. Our numerical analysis shows that the competition between the quasiperiodic disorder and inhomogeneous interaction leads to a {localization-delocalization} transition as the interaction strength is tuned from attractive to repulsive in nature. Furthermore, we analyse the combined effect of the SO and Rabi coupling strengths on the localization transition for different interaction strengths and obtain signatures of similar localization-delocalization transition as a function of SO coupling in the regime of weak interactions. We complement our numerical observation with the analytical model using the Gaussian variational approach. In the end, we show how the localization-delocalization is manifested in the quench dynamics of the condensate. Our study provides an indirect approach to achieve localization transition without tuning the quasiperiodic potential strength, but rather by tuning the inhomogeneity in the interaction. 
\end{abstract}			

\maketitle

\section{Introduction}
\label{sec1}
After the path breaking prediction of exponential localization of electronic wavefunction in random media by Anderson, the study of localization transition has attracted a great deal of attention in condensed matter~\cite{Anderson:1958, Melvin1969, TVR1979,RevTVR1985, Ferdinand2008}. Recent developments in the experimental front, such localization transitions have been observed in various artificial systems such as electromagnetic waves~\cite{Wiersma:1997, Scheffold:1999, Schwartz:2007, Aegerter:07, Topolancik:2007}, microwaves ~\cite{Dalichaouch:1991, Dembowski:1999, Chabanov:2000, Pradhan:2000}, acoustic waves~\cite{WEAVER:1990}, quantum matter waves~\cite{Lagendijk:2009, Aspect:2009}, and in the non-interacting~\cite{Billy:2008} and interacting Bose-Einstein condensates(BECs)~\cite{Roati:2008}. 

Among the various platforms, enormous progress has been made in studying localization transition in systems of BECs in random and quasiperiodic lattices due to the flexibility in controlling the geometry and interaction. In this context, more complex systems have been considered recently, such as the binary BECs~\cite{Hall:1998, Ostrovskaya:2004, Xi:2014}, spin-orbit (SO) coupled spinor BECs, etc.~\cite{lin:2011, cheng:2014, Li:2016,wang:2019, Santos:2021, Zezyulin:2022}. Theoretically, it has been demonstrated that the SO coupling influences the appearance of localization, whereas the Rabi coupling slightly promotes the delocalization~\cite{Oztas:2019}. Akin to SO coupling, interactions among atoms play a vital role in dictating the localization-delocalization phenomena in the condensate. For instance, several numerical studies show that the increase of repulsive interaction induces the localization to delocalization transitions for the condensate trapped in the disordered potential~\cite{ Sanchez:2007, Adhikari:2009, Adhikari:2010, Cheng_random:2010, Cheng:2010, Cheng:2011}.  Subsequent studies have reported the symmetry breaking of a localized binary condensate in the presence of inter-species and intra-species atomic interactions~\cite{Cheng:2010, Cheng_spat:2010}. Santos and Cardoso demonstrated that even with only one component subjected to the quasiperiodic potential, the interspecies interaction induces the localization in the other component~\cite{Santos:2021}. 

In BECs, atomic interactions play a major role in modifying the localization-delocalization phases of a condensate. Usually, in experiments, the homogeneous atomic interactions can be tuned using the external magnetic field in Feshbach resonance technique. Apart from the magnetic Feshbach resonance, Fedichev {\it et al.}~\cite{Fedichev:1996} theoretically proposed a technique to control the atomic interactions using the optical laser instead of using external magnetic field. Following their prediction, Theis {\it et al.}~\cite{Theis:2004} realized this phenomena in experiment with Rb$^{87}$ atoms which is popularly known as optical Feshbach resonance. This technique allows atomic interactions to be controlled by the detuning of the optical laser, which ultimately induces spatial inhomogeneity in the atomic interactions. The spatial variation in the atomic scattering length has become commonly referred to as a {\it collisionally inhomogeneous} environment.
 Later on, a great deal of effort has been made in exploring areas like adiabatic compression of matter waves~\cite{Theocharis:2005, Rodas-Verde:2005, Carpentier:2006, Theocharis:2006}, dynamical trapping of matter wave solutions~\cite{Garnier:2006}, soliton oscillations~\cite{Niarchou:2007, Belmonte:2008}, BECs in nonlinear optical lattices~\cite{GOLAMALI:2009}, and localization in the scalar~\cite{Cheng:2011, Sudharsan:2013} or binary BECs~\cite{Cheng:2009}. The investigations on localization in the presence of {\it collisionally inhomogeneous} environment reveal the effect of spatially dependent interaction towards localization as a result of the genesis of an effective potential whereas, without the inhomogeneity, the atomic interaction leads to delocalization of the condensate~\cite{Cheng:2009, Cheng:2011, Sudharsan:2013}.

%Another sort of technique that has been utilized to analyze the localization and delocalization of the condensate is the spatial variation of atomic scattering length leading to {\it collisionally inhomogeneous} environment~\cite{Fedichev:1996} upon tuning the spatial variation of laser field intensity with proper choice of detuning parameters. In this direction, Theis {\it et al.}~\cite{Theis:2004} experimentally generated the spatial variation of scattering length utilizing the collisional dynamics of BECs. Following this, a large amount of effort has been invested in exploring areas like adiabatic compression of matter waves~\cite{Theocharis:2005, Rodas-Verde:2005, Carpentier:2006, Theocharis:2006}, dynamical trapping of matter wave solutions~\cite{Garnier:2006}, soliton oscillations~\cite{Niarchou:2007, Belmonte:2008}, BECs in nonlinear optical lattices~\cite{GOLAMALI:2009}, and localization in the scalar~\cite{Cheng:2011, Sudharsan:2013} or binary BECs~\cite{Cheng:2009}. The investigations on localization in the presence of {\it collisionally inhomogeneous} environment reveal the effect of spatially dependent interaction towards localization as a result of the genesis of an effective potential whereas, without the inhomogeneity, the atomic interaction leads to delocalization of the condensate~\cite{Cheng:2009, Cheng:2011, Sudharsan:2013}. 

Apart from the standard delocalization-localization transition, a new class of localization transition known as the reentrant localization transition in quasiperiodic lattices has attracted a great deal of attention in recent years. While in the conventional localization transition, the states after the localization transition remain localized with an increase in the disorder in the system, in the case of reentrant localization, the localized system delocalizes again before getting localized for the second time as a function of the quasiperiodic potential strength~\cite{Roy:2021,Padhan:2022, Roy:2022,Zhu:2009}. This kind of behavior can also be described as localization-delocalization transition. Such intriguing reentrant localization has been studied extensively in recent years~\cite{roy:2020epl}, including in non-Hermitian lattices~\cite{suotang_jia}, leading to their experimental observations~\cite{Qi2023, Jiang2021, Rechtsman2024, Ali2024}.  {In that context, few theoretical studies have been reported about intermediate critical phases where extended and localized states co-exist in the presence of spin-orbit interaction and quasiperiodic disordered potential in strongly correlated system~\cite{Zhou:2013, Ray:2016,Ray:2017}.}
In the continuum systems, however, this kind of localization-delocalization-localization behavior has not been analyzed with the same rigour as those in the strongly correlated system. Although a recent study has revealed the localization-delocalization transition upon variation of SO coupling in a \textit{non-interacting} condensate trapped in a disordered quasiperiodic potential~\cite{Li:2016}. However, a complete understanding of the reentrant localization for weakly \textit{interacting} BECs is lacking.

In this work, we present an extensive numerical study that explores the impact of inhomogeneous interactions in the localization of SO coupled BECs subjected to a quasiperiodic optical lattice potential. We consider the spatial dependent interactions induced by the external optical fields to have the same form as the quasiperiodic potential with the parameters depending on the intensity of the lasers~\cite{Sakaguchi:2005, ABDULLAEV:2007}. This setting leads to the localization of the condensate with the variation of interaction strength. However, by implementing an additional phase shift $\pi / 2$ between the inhomogeneous interaction and trapping potential, we obtain the signature of localization-delocalization transition, where we find an intermediate delocalized phase sandwiched between two localized phases with the variation of the strength of the inhomogeneity of the nonlinearity. Furthermore, we complement this behavior in the localization through the Gaussian variational approach. At last, we obtain the signatures of such localization properties from the dynamical behavior of the condensate.

%\sout{ with different approaches, such as perturbing the condensate using velocity and quenching of the trapping potential.}

%, which has been systematically captured by the numerical and the Gaussian variational approach. Furthermore, the transition from delocalized to localized states can be determined using the condensate's width, form factor, and total energy. In the second part, we implement an additional phase shift $\pi / 2$ between inhomogeneous interaction and trapping potential. As a result, the competition between interaction and trapping potential leads to the reentrant feature of the localization, where an intermediate delocalized phase is sandwiched between two localized phases with the variation of nonlinearity. At last, we study the dynamical behaviour of the condensate with different approaches, such as perturbing the condensate using velocity and quenching of the trapping potential.

The paper is structured as follows. Section~\ref{sec2} introduces a mean-field theoretical model to examine SO coupled BECs in a quasiperiodic optical lattice potential with spatially varying interactions. Section~\ref{sec:1A} presents the numerical results, starting from condensate densities depending on inhomogeneous interactions and various coupling parameters. Here, we characterize the localized and delocalized states using the form factor and condensate width. Section~\ref{sec:C} explores the role of inhomogeneous interactions in demonstrating the localization-delocalization phases of the condensate. {In section~\ref{sec:III_B}, we present the Gaussian variational approach  to understand the characteristic of the localization based on the effective potential}. Section~\ref{Sec:D} investigates the dynamics of localized and delocalized condensates under velocity perturbations and trap quenching. Finally, Section~\ref{sec:IV} provides a summary and conclusion. Appendices~\ref{Appen:1} and ~\ref{appen:B} provide detailed equations of motion and discussions on variational approaches and Lagrangian details, respectively, along with numerical and variational comparison and energy analysis.

\section{Governing dynamical Mean-field equation and simulation details}
\label{sec2}
%We consider a pseudo spin-$1/2$ quasi-1D condensate, strongly trapped along the transverse direction, which can be modeled using the coupled Gross-Pitaevskii equations 
We consider a pseudo spin-$1/2$ quasi-1D condensate with strong transverse confinement, modelled using the coupled Gross-Pitaevskii equations ({GPEs}) as~\cite{li:2012,Achilleos:2013, vinay:2017}, 
\begin{subequations}
\label{eqn1}
\begin{align}
% \begin{split}
\mathrm{i}\frac{\partial \psi_{1}}{\partial t} = & \Bigg[-\frac{1}{2}\frac{\partial^2}{\partial x^2} + \left(g_{1 1}\vert \psi_{1}\vert ^2+ g_{1 2}\vert \psi_{2}\vert ^2\right)+ V(x)\Bigg]\psi_{1} \notag \\ &
- \mathrm{i} k_L\frac{\partial \psi_{1}}{\partial x} + \Omega \psi_{2}, \label{eqn1(a)} %\\ 
\end{align}
\begin{align}
\mathrm{i}\frac{\partial \psi_{2}}{\partial t} = & \Bigg[-\frac{1}{2}\frac{\partial^2}{\partial x^2} + \left(g_{2 2}\vert \psi_{2}\vert ^2 + g_{2 1}\vert \psi_{1}\vert ^2\right) + V(x)\Bigg]\psi_{2} \notag \\ &
 +\mathrm{i} k_L\frac{\partial \psi_{2}}{\partial x} + \Omega \psi_{1}, \label{eqn1(b)} 
\end{align}
\end{subequations}
where $\psi_{1}$ and $\psi_{2}$ represent the wavefunctions for pseudo spin-up and spin-down components, respectively. Here, $g_{1 1}$ and $g_{2 2}$ are the intra-species nonlinearities for spin-up and spin-down components, and $g_{1 2}$ represents the interspecies interaction strength. $k_L$ is the SO coupling strength and $\Omega$ is the Rabi coupling strength. $V(x)$ represents the trapping potential. For our studies, we consider $g_{1 1} = g_{2 2} = g$. The wave functions follow the normalization condition as 
\begin{align}
 \int_{-\infty}^{\infty} \left( \vert \psi_{1} \vert^2 + \vert \psi_{2} \vert^2 \right) dx = 2.
 \label{eqn2}
\end{align}
To obtain the non-dimensionalized Eq.~\ref{eqn1}, we consider the transverse harmonic oscillator length $a_{\perp} = \sqrt{\hbar/(m\omega_{\perp})}$ as a characteristic length scale with $\omega_{\perp}$ as the transverse harmonic trapping frequency, $\omega^{-1}$ as the timescale and $\hbar \omega_{\perp}$ as the characteristic energy scale. The interaction parameters can be defined in terms of $g = 2Na_{1 1}/a_{\perp}$, and $g_{1 2} = 2Na_{1 2}/a_{\perp}$, where $a_{1 1}$ and $a_{1 2}$ represent the intra-component and inter-component scattering lengths, respectively. The SO coupling is re-scaled as $k_L \equiv k_L' a_{\perp}$ and Rabi coupling as $\Omega \equiv \Omega' / (2\omega_{\perp})$. The wavefunction is rescaled as $\psi_{1, 2 }(x,t) \equiv \psi'_{1, 2 }(x,t)\sqrt{a_{\perp}}$. Here variables with prime represent the dimensional quantities.

To analyze the characteristics and dynamics of the localization of the condensate, we consider a quasiperiodic trapping potential $V(x)$ of the form~\cite{Li:2016}, 
%In order to analyze the characteristics and dynamics of the localization of the condensate, the trapping potential $V(x)$ is considered to be quasiperiodic which is of the form~\cite{Li:2016}, 
\begin{align}
V(x) = - V_1 \cos(k_1 x) - V_2 \cos(k_2 x), \label{eqn3} 
\end{align}
where $V_1$ and $V_2$ are the primary and secondary optical lattice amplitudes, respectively. For all calculations, we consider the primary and secondary lattice strengths as $V_1 = 1$ and $V_2 = 0.1$, respectively, with the angular wavenumber ratio as the golden ratio, i.e., $k_2/k_1 = (\sqrt{5} + 1)/ 2$. With this angular wave number, the wavelength of the laser beams will have the values as $\lambda_1 \approx 3.1415 a_{\perp}$ and $\lambda_2 \approx 1.9416 a_{\perp}$, respectively, where $a_{\perp}$ denotes the characteristic length scale approximately equal to $1.17 \mu$m. Note that in the experiment of Roati {\it et al.}~\cite{Roati:2008}, the ratio $k_2/k_1$ is $1.1971$.

To make the parameters considered in numerical simulations experimentally feasible, we consider the two hyperfine states of $^{87}$Rb atom as pseudo spin-up state $\ket{1} \equiv \ket{F = 1, m_F = 0}$, and pseudo spin-down state $\ket{2} \equiv \ket{F = 1, m_F = -1}$. Following the experiment of Lin {\it et al.}~\cite{lin:2011}, we consider $N \sim 1.8 \times 10^5$ number of atoms in a harmonic trap strongly confined along the perpendicular direction with frequency $\omega_{\perp} \approx 2\pi \times 83.66$ Hz, and along the axial direction the trapping frequency is chosen to be $\omega_{x} \approx 2\pi \times 7.07$ Hz. In the experiment, the wavelength of the Raman lasers, $\lambda_L = 804.1$nm and its geometry provide a tool for tuning the SO-coupling strength ($k_L$). The SO coupling strength $k_L$ can be tuned within the range $k'_L =[0.1 - 4]a_{\perp}$ using a pair of Raman lasers. On the other hand, the intensity of the Raman lasers is attributed to the variation of Rabi coupling frequency $\Omega$, where, in experiments, it can be adjusted within the interval $\Omega' = 2 \pi \times [105.13 - 669.28]$Hz, which is associated to the dimensionless Rabi-coupling frequency $\Omega = [0.1 - 4]$. Furthermore, by choosing the s-wave scattering length values between $a_{1 1} = a_{2 2} = [-0.0866a_0, 0.0866a_0] $ the intra-species interaction parameters can be adjusted between $g_{2 2} =g_{1 1} =[-1, 1]$. Similarly, for inter-species interaction parameters $g_{1 2}$, the scattering length $a_{1 2}$ can be varied from $-0.0866a_0$ to $0.0866 a_0$, where $a_0$ is the Bohr radius.
%{Furthermore, without an external magnetic field, the atomic background scattering lengths $a_{11}, a_{22},$ and $a_{21}, a_{12}$ are of order of $100a_0$, where $a_0$ is the Bohr's radius. For the interaction inhomogeneity, In reference \cite{Theis:2004}, authors have reported the variation of the scattering lengths between $10a_0$ to $190a_0$ by tuning the detuning of the laser. }

One of the main focuses of the present work is to show the role of inhomogeneity in the interaction, describing the localization and delocalization of the condensate. For that, here we provide a generalized expression for the spatial dependence of the atomic scattering length depending upon the intensity $I(x)$ and detuning parameter of the laser beam. In general, the spatial dependence of the scattering length can be expressed as~\cite{Fedichev:1996,Theis:2004,Sakaguchi:2005,Sudharsan:2013},
%\textcolor{red}{FROM THE FOLLOWING DISCUSSION: NOT CLEAR ABOUT THE SPATIAL DEPENDENCE OF INTERACTION}
\begin{subequations}
\label{eqn4}
\begin{align}
 a(x) = &\, a^0 + \frac{\alpha I(x)}{\gamma_s + \beta I(x)} \label{eqn4a} \\
 %a(x) = &\, a^0 + \frac{1}{2k}\frac{\Gamma_{\rm stim} \Delta}{\Delta^2 + (\frac{\gamma_{s}}{2})^2} \label{eqn4a}\\
 %a_{1 2}(x) = &\, a_{1 2}^0 + \frac{1}{2k}\frac{\Gamma_{\rm stim} \Delta}{\Delta^2 + (\frac{\gamma_{s}}{2})^2} \label{eqn4b}
 a_{1 2}(x) = &\, a_{1 2}^0 + \frac{\alpha I(x)}{\gamma_s + \beta I(x)}\label{eqn4b}
\end{align}
\end{subequations}
Here, $a^0$ and $a_{1 2}^0$ denote the intra- and interspecies s-wave scattering lengths when the optical fields are absent. $\alpha$ and $\beta$ are the constants depending upon the laser detuning~\cite{Theis:2004}. $\gamma_s$ is the spontaneous decay rate between two atomic levels. In the experiment, the laser detuning is considered to be larger than the spontaneous decay rate that minimizes the decay of atomic state, which in turn helps to realize the optical Feshbach resonance. Here, $I(x)$ is the intensity of the laser beam generated due to the stimulated transition between two atomic states~\cite{Julienne:1997, Julienne:2005, Julienne:2006}. Note that, the above formula is approximated for large detuning of the laser.
%$I(x)$ is the intensity of the laser beam originated from the stimulated transition between two levels of an atom. However, the dependence of stimulated transition rate towards intensity ($I(x)$) of the laser has been proposed theoretically in Ref.~\cite{Julienne:1997, Julienne:2005, Julienne:2006}. 

Theis {\it et al.}~\cite{Theis:2004} experimentally demonstrated that the scattering length of $^{87}$Rb condensate can be measured upon variation of laser detuning utilizing the optical Feshbach resonance technique. They estimated that the scattering length $a$ is within the range $10a_0$ to $190a_0$, considering the background scattering length value of $^{87}$Rb as $a^0 = 100a_0$ resulting in the variation of the scattering length in the range $-90a_0$ to $+90a_0$. One can tune the interaction from attractive to repulsive regions with the above technique. Recently, Carli {\it et al.}~\cite{Carli:2020} experimentally demonstrated the dynamical evolution of {BECs} in a collisionally inhomogeneous environment. They observed various dynamical phases of expanding condensates in a quasi-1D geometry, including the formation, decay, and cascading of soliton-like phases. Their investigation of interaction inhomogeneity opens new avenues for exploring various phenomena in BECs. 

The spatial dependence of the scattering length $a(x)$ can be achieved using the space dependence intensity, $I(x)$ of the laser beam as described in Eq.~(\ref{eqn4a}) and (\ref{eqn4b}). Without loss of generality, it is reasonable to consider the nonlinearity proportional to the optical lattice potential $V(x)$, as it is created using the laser field intensity \cite{Sakaguchi:2005, Cheng:2009, Cheng:2011}. In this work, we assume $a_{1 1}(x) = a_{2 2}(x) = a(x)$. With this consideration, $g_{1 1}(x)$, $g_{2 2}(x)$ and $g_{1 2}(x)$ can be written as, 
\begin{subequations}
\label{eqn5}
\begin{align}
g_{1 1}(x) = &\, \epsilon_{0} + \epsilon V(x),\label{eqn5a} \\
g_{2 2}(x) = &\, g_{1 1}(x),\label{eqn5b} \\ 
g_{1 2}(x) = &\, \eta_{0} + \eta V(x)\label{eqn5c}
\end{align}
\end{subequations}
Here, $\epsilon_0$ and $\eta_0$ are the homogeneous interactions associated with the background scattering length. Similarly, $\epsilon$ and $\eta$ denote the inhomogeneity interaction strength. Using the optical Feshbach resonance as the inhomogeneous interaction could be tuned independently compared to the homogeneous counterpart~\cite{Theis:2004}. Therefore, for our analysis, we consider the homogeneous interaction as $\epsilon_0 = \eta_0 = 0$ to focus totally on the effect of the inhomogeneous part of the interaction on the localization-delocalization~\cite{Sudharsan:2013, Adhikari:2010}. 

%Without losing the generality we set $\epsilon_0 = \eta_0 = 0$ in all our investigations~\cite{Sudharsan:2013, Adhikari:2010}.}

%{ At this point, it is worth noting a crucial condition where the stimulated decay rate is significantly lower than the natural decay rate between two-level atoms to control the atomic interactions via optical Feshbach resonance~\cite{Theis:2004}. However, the larger value of laser detuning compared to the spontaneous decay rate results in a greater contribution of the inhomogeneous interaction relative to the background scattering length. Therefore, to focus on studying the impact of inhomogeneity without losing generality, we set $\epsilon_0 = \eta_0 = 0$ in all our investigations~\cite{Sudharsan:2013, Adhikari:2010}.} 
%to investigate the sole impact of inhomogeneity, we choose $\epsilon_0 = \eta_0 = 0$ in all of our studies.

%In the second part of our study, to investigate localization arising from the interplay between the trapping potential and nonlinearity,

We also consider a scenario where the trapping potential (\ref{eqn3}) and spatially dependent nonlinear interactions (\ref{eqn5a})-(\ref{eqn5c}) have a phase difference $\theta_1, \theta_2$. With this the nonlinear interactions take the form as:
\begin{subequations}
\begin{align}
 g_{1 1}(x) = - \epsilon \sum_{l=1}^{2}V_l \cos \left(k_l x - \theta_1 \right) \label{eqn6a} \\
% \end{align}
% \begin{align}
 g_{1 2}(x) = - \eta \sum_{l=1}^{2}V_l \cos \left(k_l x + \theta_2 \right)\label{eqn6b}
\end{align} 
\end{subequations}
{Here, $V_l$ denotes the optical lattice amplitude, identical to $V_1$ and $V_2$ as described in Eq.~\ref{eqn3}. Similarly, $k_l$ represents the angular wavenumber corresponding to the wavelength of the laser beam. Both $V_l$, and $k_l$ are kept same as the quasiperiodic potential (Eq.~\ref{eqn3}).}
{In our case, we consider two scenarios: (i) $\theta_1 = \theta_2 = \theta = 0$, and (i) $\theta_1 = \theta_2 = \theta = \pi/2$. For the first case, the spatial inhomogeneity is the same as the quasiperiodic potential with $\theta = 0$. In contrast, the $\theta = \pi/2$ introduces a phase shift between interaction inhomogeneity and quasiperiodic potential. To further generalize our analysis, we also explore cases with unequal phase shifts i.e., $\theta_1 \neq \theta_2$, between the interaction and the quasiperiodic potential as discussed in \ref{appen:B}.
}
%\sout{As a second case, we have chosen a spatial inhomogeneity in the interspecies interaction to be the same as that of the quasiperiodic potential but with a phase difference $\theta_1 = \theta_2 = \theta = \pi/2$.}

The dynamical equations [Eqs.~\ref{eqn1}(a) and \ref{eqn1}(b)] are solved using the split-step Crank-Nicholson method~\cite{Murug:2009, Young:2016, RAVISANKAR:2021} with the box size as $[-102.4:102.4]$, space step $\Delta x = 0.025$ and time step $\Delta t = 10^{-4}$. For all simulations, we consider the Gaussian wavefunction centered at zero as an initial state for both components and also assume the antisymmetric condition $\psi_{1}(x) = - \psi_{2}(-x)$ between the components as it represents the ground state of the Hamiltonian with lowest energy~\cite{Adhikari:2010,Ravisankar:2020}. While solving the dynamical equations, we employ an imaginary-time scheme to obtain the ground state and a real-time propagation scheme to investigate the condensate dynamics. 

%%%%%%%%%%%%%%%%%%%
\section{ Results}
%%%%%%%%%%%%%%%%%%%%%%%%%%%%%%%%%%%%%%%%%%%%%%%%%%%%

%%%%%%%%%%%%%%%%%%%%%%%%%%%%%%%%%%
\begin{figure}[!htb]
\centering
\includegraphics[width=\linewidth]{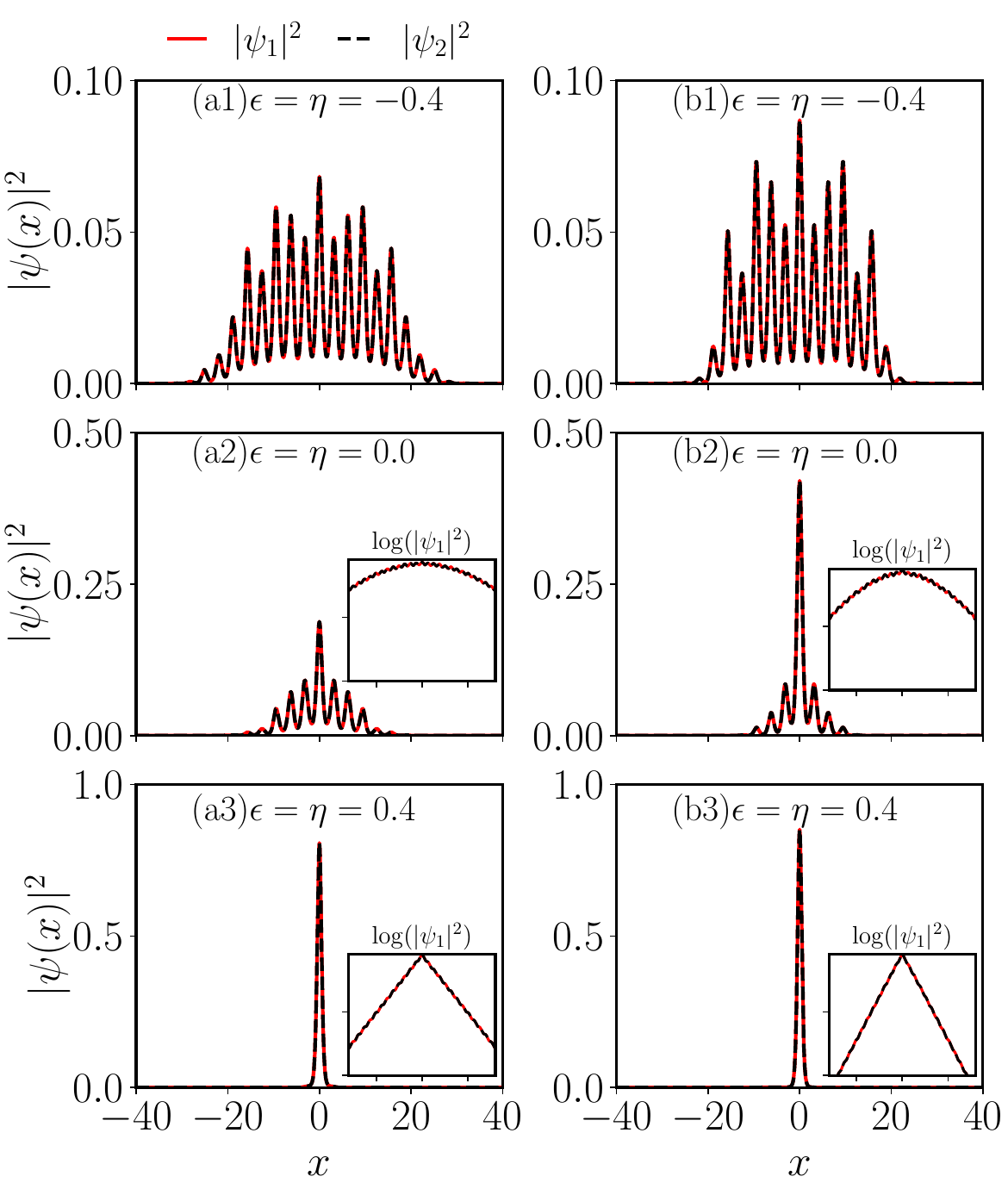}
\caption{Density profile of $|\psi_1|^2$ (red line) and $|\psi_2|^2$ (dashed black line) components for different inhomogeneity parameter $\epsilon=\eta=-0.4$ (a1,b1), $\epsilon=\eta=0$ (a2,b2), and $\epsilon=\eta=0.4$ (a3,b3) at $k_L=0$, $\Omega =0.0$ (left column) and $k_L=0.6, \Omega = 1$ (right panel). Insets in [(a2, a3, b2, b3)] show the spatial variation of density profile in semilog scale that explicitly displays the exponential tail of the localized condensate in (a3) and (b3).}
\label{den-diff-kL}
\end{figure}
%%%%%%%%%%%%%%%%%%%%%%%%%%%%%%%%%%%%%%%%%%%%%%%%%%%%

In this section, we present the effect of the inhomogeneous interaction and the Rabi and SO coupling on the spatial profile of the ground state obtained using the imaginary time propagation. First, we focus on analyzing the impact of the inhomogeneity on the localization of the ground state. Following this, we investigate the effect of phase shift between the interaction and potential in bringing the interesting interplay between the localized and delocalized state that finally leads to the manifestation of the localization-delocalization features of the localization with change in the interaction inhomogeneity. We complement our numerical observation by employing the Gaussian variational approach. Finally, we present a detailed nature of the dynamics of the localized and delocalized state induced by several means, namely, by giving small velocity perturbations and quenching of the potential.

%%%%%%%%%%%%%%%%%%%%%%%%%%%%%%%%%%%%%%%%%%%%%%%%%%%%%%%%
\subsection{Effect of interaction inhomogeneity on the Localizatio-delocalization transition of the condensate}
\label{sec:1A}

In a recent work~\cite{Li:2016} Li {\it et al.} demonstrated that the condensate remains localized for a higher ratio of secondary to the primary optical lattice strength ($V_2/V_1$) for all the range of SO and Rabi coupling parameters. However, for small $V_2/V_1 (\sim 0.1)$, the condensate exhibits a delocalized nature, which undergoes a transition to the localized state upon either increasing the Rabi coupling for a fixed SO coupling or increasing the SO coupling for a fixed Rabi coupling. %
%%%%%%%%%%%%%%%%%%%%%%%%%%%%%%%%%%%%%%%%%%%%%%%%%%%%%%%%%%%%%%%%%%
\begin{figure}[!htp]
\centering
\includegraphics[width=\linewidth]{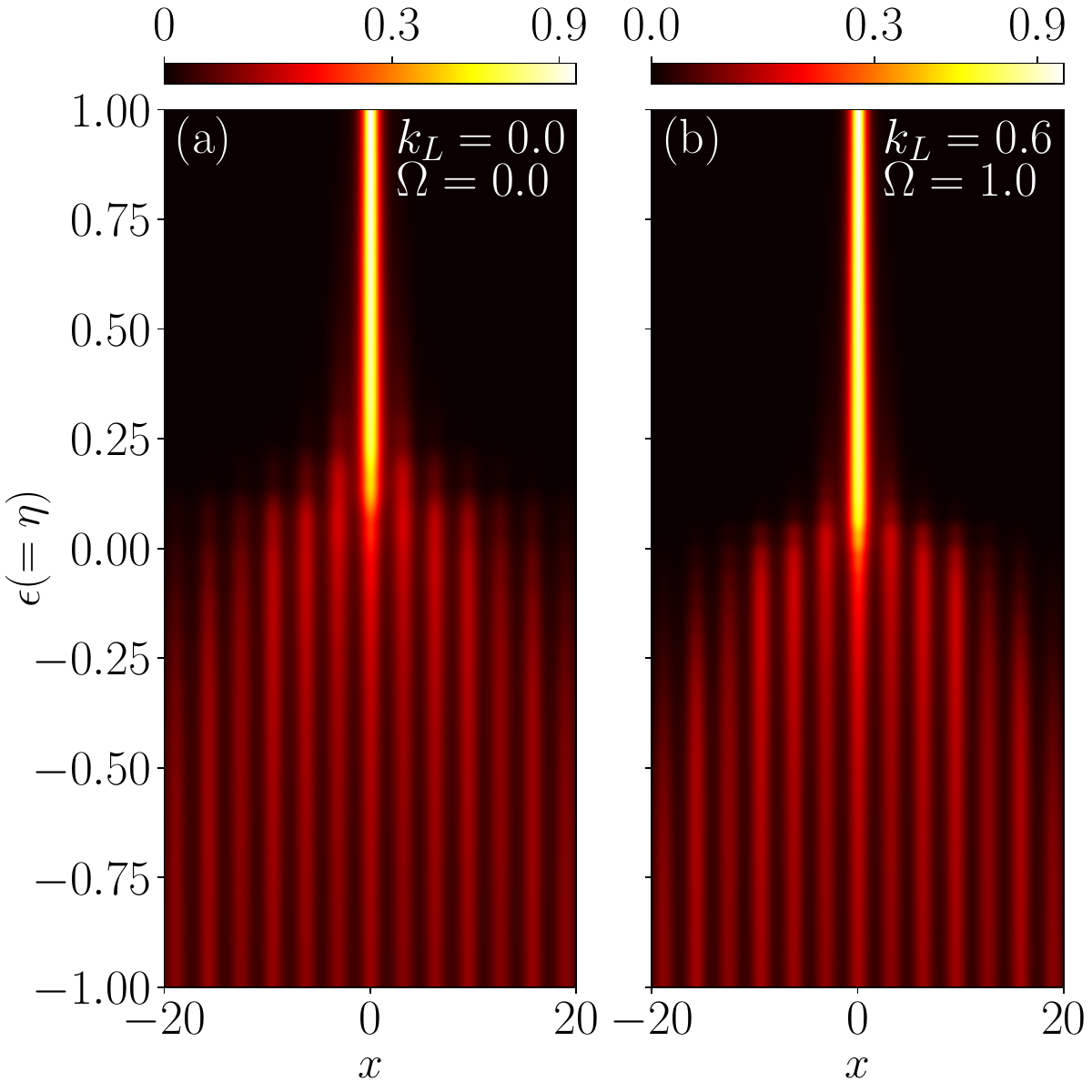}
\caption{Pseudo color representation of the spatial variation of ground state density of the spin-up component ($\lvert \psi_1 \rvert^2$) with the inhomogeneity parameter $\epsilon=\eta$ for (a) $k_L=0$, $\Omega = 0$ and (b) $k_L=0.6$, $\Omega = 1.0$. For the zero coupling parameter, the condensate exhibits a delocalized state for $\epsilon=\eta\lesssim 0.2$ and a localized state for $\epsilon=\eta> 0.2$. However, for finite coupling parameter ($\Omega=1$, $k_L=0.6$) the localization state is realized for $\epsilon=\eta\gtrsim 0$.}
\label{den-var-epsilon-eta}
\end{figure}
%%%%%%%%%%%%%%%%%%%%%%%%%%%%%%%%%%%%%%%%%%%%%%%%%%%%%%%%%%%%
While in the former case, the condensate remains delocalized with small value of SO coupling strength, whereas, for moderate SO coupling strength the condensate remain localized at high Rabi coupling. In the latter case, it shows the localization-delocalization behavior with the change in the SO coupling for small Rabi coupling.
%\sout{While in the former case, it remains localized for high Rabi coupling, in the latter case, it shows the reentrant nature of the localization with the change in the SO coupling.} 
{In this work, we aim to explore the effect of interaction inhomogeneity alongside  the SO and Rabi coupling parameters on the state of the condensate, which is initially delocalized when no inhomogeneity or coupling parameters are included. In the following, we will examine how these factors, namely, inhomogeneity and the coupling parameters overall affect the condensate's localization by  fixing the lattice strength ratio as $V_2/V_1=0.1$.} 
%{In this study, we aim to analyze the effect of inhomogeneous interactions along with the impact of SO and Rabi coupling, starting from a delocalized condensate. To investigate the influence of these interactions and coupling parameters, we fix the lattice strength at $ V_2/V_1 = 0.1 $ for all cases.}

%In a recent work~\cite{Li:2016} Li {\it et al.} demonstrated that the condensate remains localized for higher ratio of secondary to the primary optical lattice strength ($V_2/V_1$) for all the range of SO and Rabi coupling parameters. However, for small $V_2/V_1(\sim 0.1)$ the condensate exhibits delocalized nature which undergoes a transition to the localized state upon either increasing the Rabi coupling for a fixed SO coupling or increasing the SO coupling for a fixed Rabi coupling. While for the former case it remains in the localized state for high Rabi coupling, in the later case it shows reentrant nature of the localization with the change in the SO coupling. In this work our aim is to analyze the effect of interaction inhomogeneity in tandem with the SO and Rabi coupling parameters for the state which is delocalized in the absence of inhomogeneity and coupling parameters. In the following, we investigate the effect of inhomogeneity and coupling parameters on the localization state of the condensate by fixing $V_2/V_1=0.1$. 

In Fig.~\ref{den-diff-kL}, we show the spatial profile of the density for different sets of spatial inhomogeneity parameters by considering the coupling parameters as $\Omega=0$, $k_L=0$ (left column) and $\Omega=1$, $k_L=0.6$ (right column). In the absence of the SO and Rabi coupling, the condensate density profile exhibits a delocalized nature for zero ($\epsilon=\eta=0$) and attractive inhomogeneity ($\epsilon=\eta=-0.4$) as depicted in the Figs.~\ref{den-diff-kL}(a2) and (a1), respectively. The condensate starts showing a tendency to localization for the repulsive inhomogeneous interaction, quite evident with the exponential tail as shown in the inset of Fig.~\ref{den-diff-kL}(a3). As we consider the SO and the Rabi coupling to a finite value, we find that the spatial spread of the condensate becomes narrower compared to those with the coupling parameters for the attractive and zero inhomogeneity parameters, which is quite evident from the Fig.~\ref{den-diff-kL}(b1) and (b2), respectively. Here, it is interesting to note that even for repulsive inhomogeneity, the condensate exhibits the localization [Fig.~\ref{den-diff-kL}(b3)] contrary to the situation when inhomogeneity is absent. Note that, in the absence of inhomogeneity, even a small repulsive nonlinear interaction destroys the localization of the condensate~\cite{Sarkar:2023}. 

%{Note that as we use the equal interspecies interaction the density of the spin-up and spin-down component remains equal. Therefore, for the sake of brevity we will show the density of spin-up component for all later part of the discussion. }

%%%%%%%%%%%%%%%%%%%%%%%%%%%%%%%%%%%%%%%%%%%%
 \begin{figure}[!htp]
 \centering
 \includegraphics[width=\linewidth]{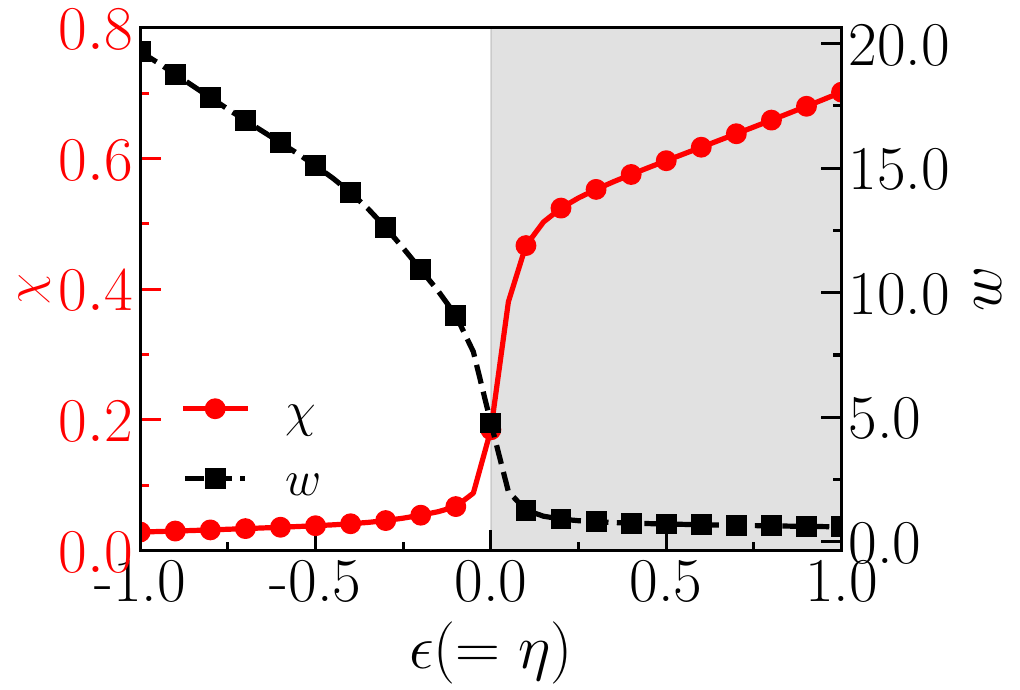}
 \caption{The form-factor $\chi$ (red circle marked solid line) and the width $w$ (black square marked dashed line) of the condensate as a function of inhomogeneity parameter $\epsilon (=\eta)$ for $k_L = 0.6$ and $\Omega = 1.0$. The increase in $\chi$ is accompanied by a decrease in the width of the condensate, which starts beyond the critical inhomogeneity parameter $\epsilon(=\eta) \geq 0.1$, indicating a transition from the delocalized to localized state. The localized range is shown with the grey-shaded area. Here, the left and right vertical axis shows the variation of $\chi$ and $w$, respectively.
 }
 \label{fig:xhi-width-compare}
 \end{figure}
%%%%%%%%%%%%%%%%%%%%%%%%%%%%%%%%%%%%%%%%%%%%%%%%

To investigate the effect of inhomogeneity in a more systematic way in Fig.~\ref{den-var-epsilon-eta}, we show the pseudocolour density profile of the condensate with a continuous variation of the inhomogeneity parameter for zero [Fig.~\ref{den-var-epsilon-eta}(a)] and finite couplings, $\Omega=1$ and $k_L=0.6$ [Fig.~\ref{den-var-epsilon-eta}(b)]. {For this case, we find that the condensate remains delocalized for entire range of the attractive interaction inhomogeneity as well as for small repulsive interaction inhomogeneity. However, we witness localization for larger repulsive inhomogeneity interaction.} The overall features indicate that for zero coupling parameters ($\Omega=k_L=0$), the density gets localized for $\epsilon(=\eta) \gtrsim 0.2$. However, with $k_L = 0.6$ and $\Omega = 1$, the condensate display localization even without interaction inhomogeneity ($\epsilon(=\eta) \gtrsim 0$). This particular feature suggests an interplay between interaction inhomogeneity and the coupling parameters towards the localization of the condensate. On the one hand, while the attractive inhomogeneity tends to delocalize the condensate, the coupling parameters try to restore the localization. To shed more light on this aspect, in the following, we will present a systematic analysis of the role of these competing factors on delocalization, which will be connected further to bring the localization-delocalization behavior of the condensate in the similar line observed in the many body system~\cite{Roy:2021}. 
\begin{figure}[!htp]
\centering
\includegraphics[width=\linewidth]{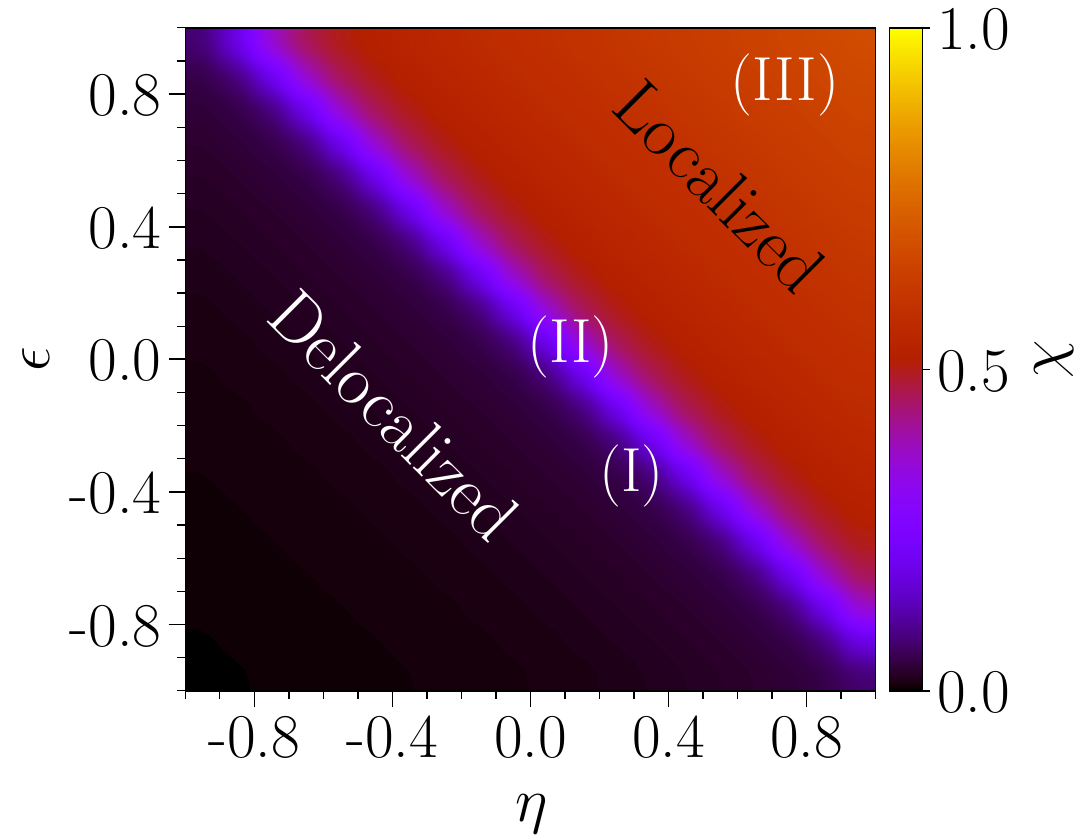}
\caption{Pseudo color representation of the form-factor ($\chi$) in the $\epsilon-\eta$ plane for $k_L = 0.6$ and $\Omega = 1.0$. For $\chi \gtrsim 0.5$, the condensate is characterized in the localized state, while with $\chi < 0.5$, it is termed as delocalized. Three distinct regions, I, II, and III, represent the delocalized, intermediate localized and localized states.}
\label{fig:phase-epsilon-eta-gamma-0p6}
\end{figure}%
%%%%%%%%%%%%%%%%%%%%%%%%%%%%%%%%%%%%%%%%%%%%%%%%%%%%%%%%%%%%%

%%%%%%%%%%%%%%%%%%%%%%%%%%%%%%%%%%%%%%%%%%%%%%%%%%%%%%%%%%%%%%%
\begin{figure*}[!htb]
 \centering
 \includegraphics[width=\linewidth]{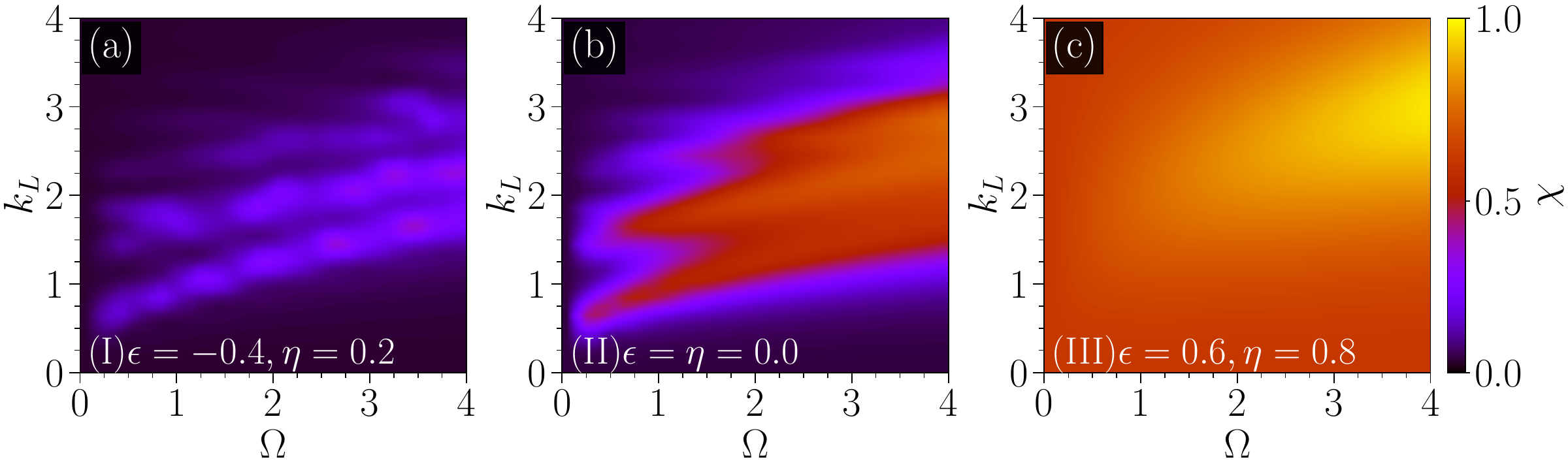}
 \caption{Pseudo color representation of the form factor $\chi$ in $k_L -\Omega$ plane in three distinct regions of the inhomogeneity parameters as indicated in the Fig.~\ref{fig:phase-epsilon-eta-gamma-0p6}: (a) for $\epsilon = -0.4$, and $\eta = 0.2$ in the region I, (b) for $\epsilon = \eta = 0.0$ in region II, and (c) for $\epsilon = 0.6, \eta = 0.8$ in region III. Regions I and III remain in the delocalized and localized states for all ranges of $\Omega$ and $k_L$. Region II exhibits the localization-delocalization transition upon tuning the SO and Rabi coupling.}
 \label{fig:phase-omega-kL}
\end{figure*}%
%%%%%%%%%%%%%%%%%%%%%%%%%%%%%%%%%%%%%%%%%%%%%%%%%%%%%%%%%%%%%

The extent of localization and delocalization of the condensate has been characterized using two quantities, namely, the integral form factor ($\chi$) and the width ($w$) of the condensate, which are defined as~\cite{Li:2016,Roy:2021},
%In order to get a comprehensive picture of the localization and delocalization phases of the system, we introduce two quantities to characterize them i.e. integral form factor and width which can be estimated after obtaining the ground state of the condensate of the form,
\begin{subequations}
\begin{align}
 \chi_j & = \frac{1}{N_j^2}\int \lvert \psi_j\rvert ^4 dx , \label{eqn11a}\\
 w_j^2 & = 2\int (x-x_{m})^2 \vert \psi_j\vert ^2 dx , \label{eqn11b} 
\end{align}
\end{subequations}
where $N_j = \int_{-\infty}^{\infty} \vert \psi_j(x)\vert ^2 dx$ represents the population of the $j^{\rm th}$ component of the condensate with $j \in \{1, 2\}$ and $x_m$ is the center of mass of the condensate. Since we choose equal number of atoms in each component, the condensate density remains same in both components, and hence, the form factor and width can be considered as $\chi_1 = \chi_2 \equiv \chi$ and $w_1 = w_2 \equiv w$, respectively. Consequently, we show the results pertaining solely to the spin-up components only. {Please note that in general the state of the condensate is categorized broadly in localized state (for $\chi \gtrsim 0.5$) and delocalized state (for $\chi\lesssim 0.5$)~\cite{Li:2016}. However, here based on the real space profile of the condensate which appears to span over a very localized region we have introduced the three state of the condensate localization.} We set the criteria that the condensate can be characterized as in the localized state if $\chi \geq 0.5$. Otherwise, for the lower form factor ($0.3\lesssim \chi \lesssim 0.5$), the condensate is in the intermediate localized state and the delocalized state for ($\chi \lesssim 0.3$).

In Fig.~\ref{fig:xhi-width-compare}, we illustrate the variation of the form factor (in the left vertical axis) and width (in the right vertical axis) of the condensate with respect to the change in the interaction inhomogeneity considering the equal inter- and intra- interaction ($\epsilon=\eta$). In accordance with the above discussion, we find that the $\chi$ remains close to zero for attractive interaction inhomogeneity ($\epsilon=\eta <0$), indicating the delocalized state of the condensate and increases to $\chi \sim 0.5$ as soon as inhomogeneity gets tuned towards the repulsive region ($\epsilon=\eta\gtrsim 0$). This trend further continues upon increasing the interaction inhomogeneity. Overall, we find that the slightly repulsive nature of the interaction inhomogeneity is good enough to localize the condensate. This feature of $\chi$ in characterizing the localization of the condensate is very well complemented by the variation of the width ($w$) of the condensate as shown using the black dashed squared line in Fig.~\ref{fig:xhi-width-compare}. The width remains high ($w\gtrsim 5$) until ($\epsilon=\eta<0$), for which the condensate remains delocalized. In the case of the localized condensate for repulsive interactions, the condensate width becomes nearly equal to one.

To understand the effect of the interaction inhomogeneity on the localization of the condensate, we perform an extensive simulation for different ranges of interaction inhomogeneity ($\lvert\epsilon\rvert = \lvert\eta\rvert <1$) by keeping the SO and Rabi coupling fixed at $k_L=0.6$ and $\Omega=1$. In Fig.~\ref{fig:phase-epsilon-eta-gamma-0p6}, we present the pseudo color of form-factor $\chi$ in the $\epsilon - \eta$ plane. Here, we have used the criteria, $\chi \gtrsim 0.5$, to designate the state of the condensate as a localized state. When both intraspecies ($\epsilon$) and interspecies ($\eta$) are positive, the condensate remains in the localized state with $\chi \gtrsim 0.6$. However, if both interactions are attractive, the form factor remains in the range $0.03 \leq \chi \leq 0.1$, indicating the delocalized state of the condensate.
%To understand the effect of the interaction inhomogeneity on the localization of the condensate, we perform an extensive simulation for different range of interaction inhomogeneity ($\lvert\epsilon\rvert = \lvert\eta\rvert <1$) by keeping the SO and Rabi coupling fixed at $k_L=0.6$ and $\Omega=1$. In Fig.~\ref{fig:phase-epsilon-eta-gamma-0p6}, we present the pseudo color of form-factor $\chi$ in the $\epsilon - \eta$ plane. Here we have used the criteria, $\chi \gtrsim 0.5$, to designate the state of the condensate as a localized state. We find that when both intraspecies ($\epsilon$) and interspecies ($\eta$) are positive the condensate remains in the localized sate with $\chi \gtrsim 0.6$. However, in case both the interactions are attractive the form factor remains in the range of $0.03 \leq \chi \leq 0.1$, indicating the delocalized state of the condensate. 

Next, we present the effect of SO ($k_L$) and Rabi ($\Omega$) couplings on the localization and delocalization of the condensate with spatial dependent interactions. For this analysis, we select three sets of parameters $\epsilon$, and $\eta$ of Fig.~\ref{fig:phase-epsilon-eta-gamma-0p6}. These parameters are associated with three different regions: region I with $\epsilon = -0.4$, $\eta = 0.2$, region II with $\epsilon = \eta = 0.0$, and region III with $\epsilon = 0.6$, $\eta = 0.8$, as marked in Fig.~\ref{fig:phase-epsilon-eta-gamma-0p6}. 

In Fig.~\ref{fig:phase-omega-kL}(a) we show the phase diagram in $k_L - \Omega$ plane for $\epsilon = -0.4$, and $\eta = 0.2$. We notice that the form factor $\chi$ remains within the range $0 \lesssim \chi \lesssim 0.4$ for all ranges $k_L$ and $\Omega$, revealing that the condensate still associates with the delocalized state. Fig.~\ref{fig:phase-omega-kL}(b) illustrates the $\chi$ in the $\Omega-k_L$ plane for the region II ($\epsilon=\eta=0$) located at the phase boundary between the delocalized and localized phases in the $\epsilon-\eta$ plane (see Fig.~\ref{fig:phase-epsilon-eta-gamma-0p6}). Furthermore, for this case, $\chi$ lies within the range $0.1 \lesssim \chi \lesssim 0.75$, implying a transition from delocalized to the localized state of the condensate. We further find the localization-delocalization transition as SO coupling is increased in the range $0 \leq k_L \leq 4$ while keeping $\Omega$ fixed at a low value ($\Omega \sim 1$). Similar localization-delocalization phase transition also have been reported earlier by Li {\it et al.}~\cite{Li:2016} by tuning the $k_L$ for non-interacting condensate. After the comprehensive picture of the condensate phase in $\Omega-k_L$ plane for the non-interacting case, next, we consider repulsive inhomogeneous interactions with $\epsilon = 0.6$, $\eta = 0.8$ marked as region III in Fig.~\ref{fig:phase-epsilon-eta-gamma-0p6}. For this case, the $\chi$ value lies within $0.60 \lesssim \chi \lesssim 1.0$, suggesting a complete localization of the condensate as shown in Fig.~\ref{fig:phase-omega-kL}(c). Interestingly, we find that the condensate becomes strongly localized ($\chi\sim 1$) for high SO and Rabi coupling parameters, even when the interactions in the condensate are repulsive in contrast to the homogeneous interaction where the repulsive interaction weakens the localization~\cite{Sarkar:2023}. {In the later part of the paper, we will relate the enhancement of the localization, which occurs with an increase in the repulsive interactions in the presence of spatial inhomogeneity, to the growing nature of the depth of the effective potential of the condensate, as derived using the variational approach.}
\begin{figure}
 \centering
 \includegraphics[width=\linewidth]{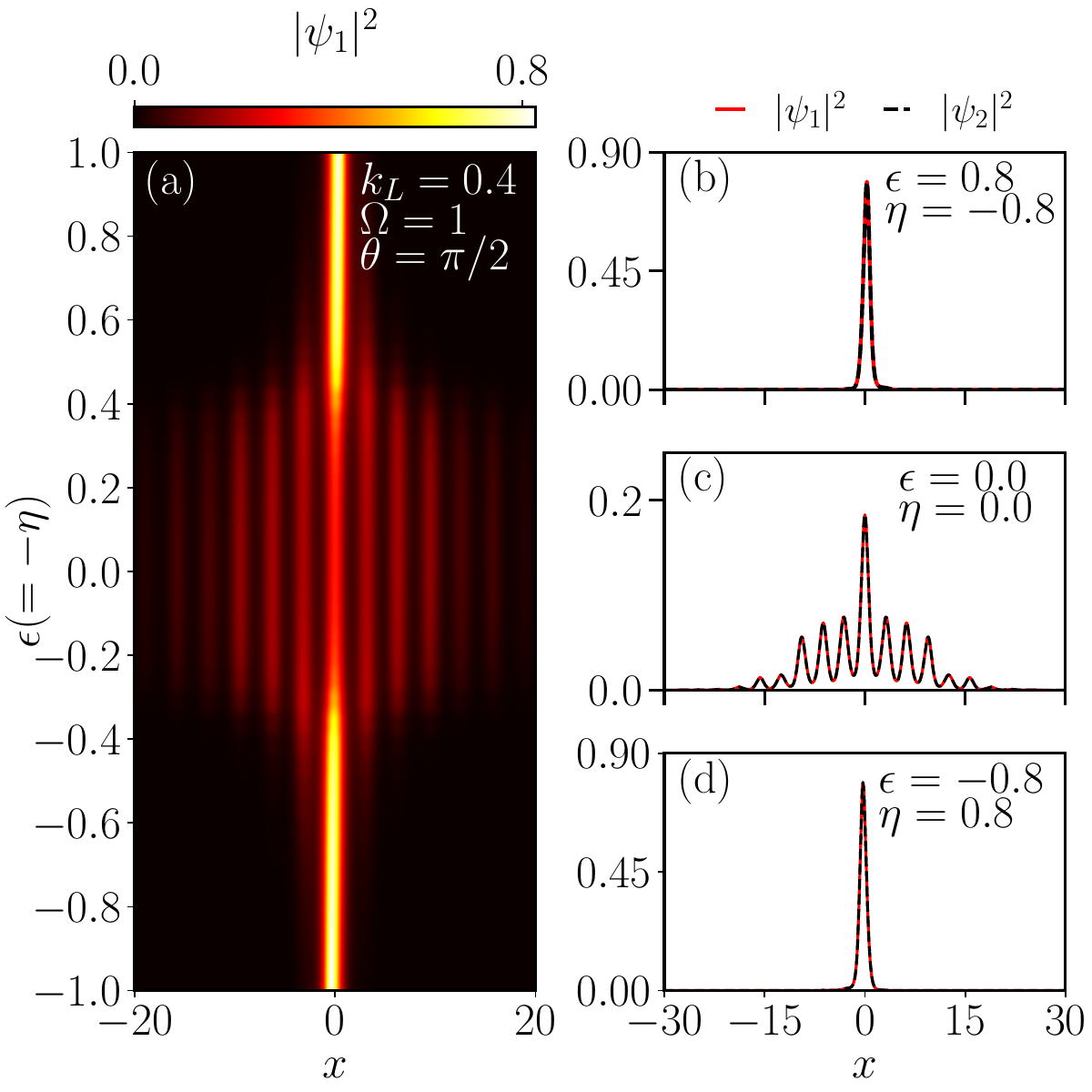}
 \caption{Effect of inhomogeneity on the density profile of the condensate for $\pi/2$ phase difference between the nonlinearity and potential with $k_L = 0.4$ and $\Omega =1.0$. (a) Pseudo color representation of density spatial profile with respect to the inhomogeneous interaction parameter $\epsilon(=-\eta)$. (b)-(d) spatial density profile for $\epsilon(=-\eta) = 0.8$ (localized), $\epsilon(=-\eta) = 0$ (delocalized), and (d)$\epsilon(=-\eta) = -0.8$ (localized). The condensate undergoes localization-delocalization-localization phases upon increasing the strength of the inhomogeneity from the attractive  to repulsive region.}
 \label{fig:density-var-epsilon}
 \end{figure}
 %%%%%%%%%%%%%%%%%%%%%%%%%%%%%%%%%%%%%%%%%
\subsection{Localization-delocalization transtion in presence of collisional inhomogeneities}
\label{sec:C}
 
In the previous section, we investigated the effect of spatially inhomogeneous interaction for the case when the spatial inhomogeneity in the nonlinear interaction is similar to those of the bichromatic lattice potential. Next, we consider the situation when there happens to be $\pi/2$ phase difference between the trapping potential and inhomogeneous interactions as given in Eqs.~(\ref{eqn6a}) and (\ref{eqn6b}).

In Fig.~\ref{fig:density-var-epsilon}(a), we present the variation of the condensate density with the inhomogeneous interaction strength $\epsilon$ and $\eta$ by assuming $\epsilon=-\eta$. We find that the condensate exhibits localization for $-0.4 \gtrsim \epsilon(=-\eta) \gtrsim -1.0$, $0.4 \lesssim \epsilon(=-\eta) \lesssim 1.0$ while, it displays delocalized nature $-0.4 \lesssim \epsilon(=-\eta) \lesssim 0.4$ as the coupling parameters are fixed to $k_L=0.4$ and $\Omega=1$. In Figs.~\ref{fig:density-var-epsilon}(b)-(d), we show the density profile for three sets of parameters $\epsilon = -\eta = 0.8$, $\epsilon = -\eta = 0$, and $\epsilon = -\eta = 0.8$, respectively. In the case of Fig. \ref{fig:density-var-epsilon}(b), the condensate is localized at $x_0 \approx 0.3$, whereas in Fig.\ref{fig:density-var-epsilon}(d), the condensate is localized at $x_0 \approx -0.3$. The shifting of the center of mass of the condensate is explained further using the effective potential using the variational approach in Sec.~\ref{sec:III_B}. In contrast, Fig.~\ref{fig:density-var-epsilon}(c) illustrates a completely delocalized phase of the condensate in the absence of inhomogeneity.
%In Fig.~\ref{fig:density-var-epsilon}(a), we present the variation of the condensate density with the inhomogeneous interaction strength $\epsilon$ and $\eta$ by assuming $\epsilon=-\eta$. We find that the condensate exhibits localization for $-0.4 \gtrsim (\epsilon,-\eta) \gtrsim 0.4$, $-0.4 \lesssim \epsilon(=-\eta) \lesssim 0.4$ while, it displays delocalized nature $-0.4 \lesssim \epsilon(=-\eta) \lesssim 0.4$ as the coupling parameters are fixed to $k_L=0.4$ and $\Omega=1$. In Figs.~\ref{fig:density-var-epsilon}(b-d), we show the density profile for three sets of parameters $\epsilon = -\eta = 0.8$, $\epsilon = -\eta = 0$, and $\epsilon = -\eta = 0.8$ respectively. In case of figure (b), the condensate localized at $x_0 \approx 0.3$, whereas, the condensate in figure (d), localized at $x_0 \approx -0.3$. The shifting of the center of mass of the condensate is explained further using the effective potential using variational approach in Sec.~\ref{sec:III_B}. In contrast to that, figure (c) illustrates a complete delocalized phase of the condensate in absence of inhomogeneity.
%%%%%%%%%%%%%%%%%%%%%%%%%%%%%%%%%%%%%%%%%%%%%%%%%%%%%%%%%%%%%%
\begin{figure}[!htp]
 \centering
 \includegraphics[width=\linewidth]{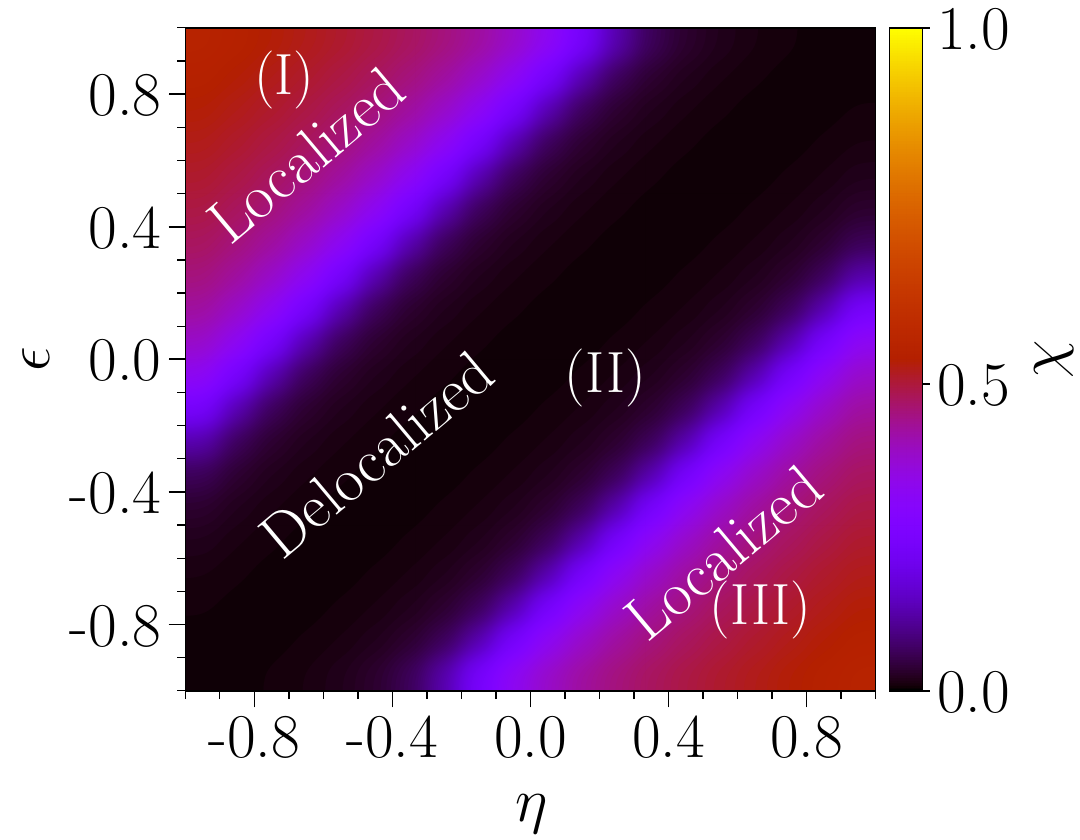}
 \caption{Pseudo color representation of the form factor $\chi$ in the $\epsilon-\eta$ plane showing localization-delocalization-localization feature of the condensate for $\pi/2$ phase between the inhomogeneity and potential. The parameters are $k_L = 0.4$, $\theta = \pi/2$, and $\Omega = 1.0$. Localized ($\chi\gtrsim 0.5$), intermediate localized ($0.3<\chi\lesssim 0.5$) and delocalized ($\chi\lesssim 0.3$) have been observed. I, II, and III denote the three distinct regimes in the inhomogeneous parameters space chosen to investigate the effect of $\Omega$ and $k_L$ on the localization and delocalization.}
\label{fig:phase-ep-eta-pib2}
\end{figure}
%%%%%%%%%%%%%%%%%%%%%%%%%%%%%%%%%%%%%%%%%%%%%%%%
\begin{figure}
 \centering
 \includegraphics[width=\linewidth]{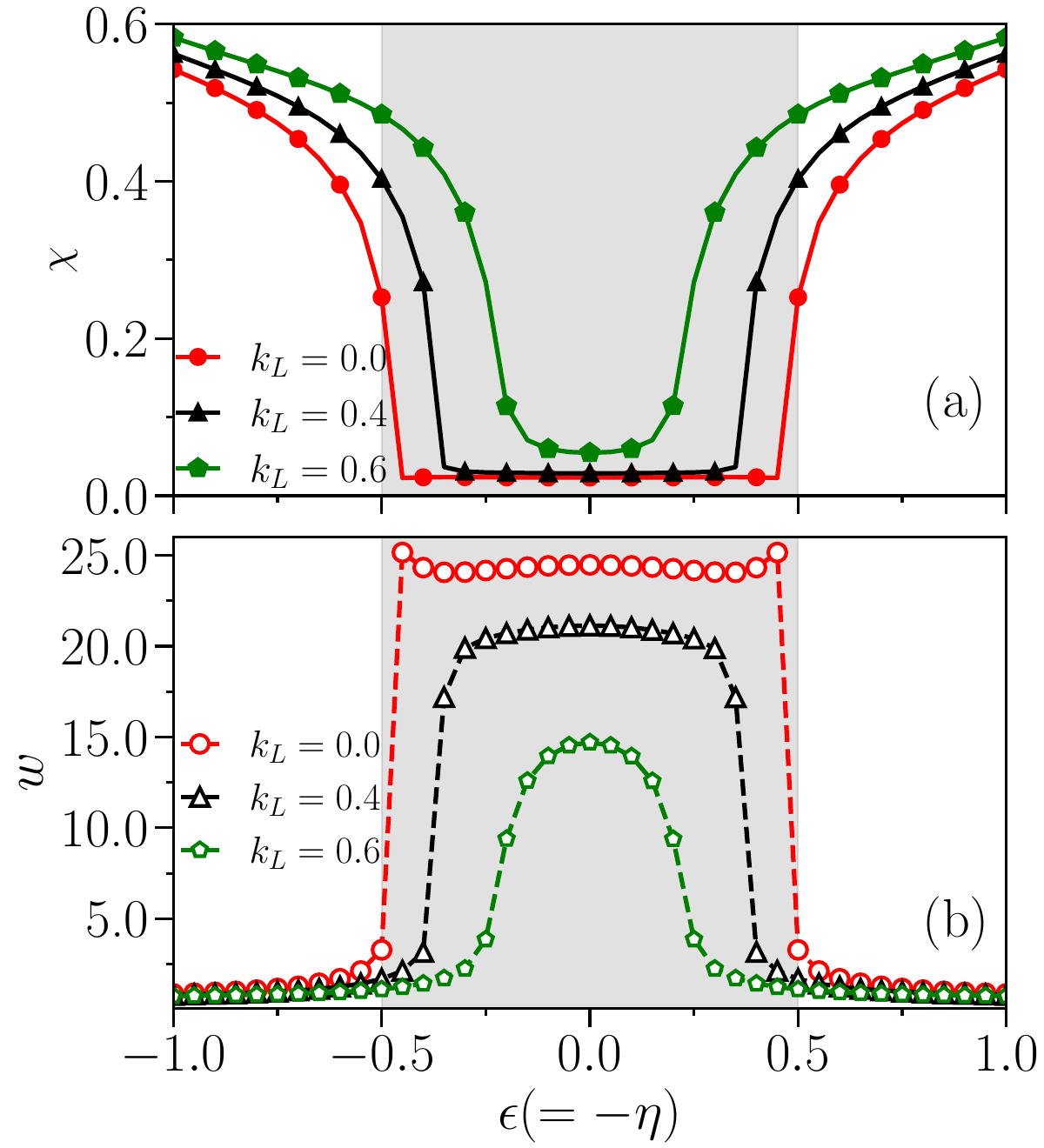}
 \caption{Variation of (a) form-factor ($\chi$) and (b) width ($w$) of the condensate as a function of the interaction inhomogeneity parameter $\epsilon = -\eta$ for $k_L = 0$ (red circles), $k_L = 0.4$ (black triangle), and $k_L = 0.6$ (green pentagons) by fixing $\Omega=1$. Increasing $k_L$ reduces the range of inhomogeneity parameters in which delocalization is observed.} 
 %range of $-0.5 \leq \epsilon(=-\eta) \leq 0.5$, as evidenced by the form-factor in (a) and also complemented by the condensate width in (b).}
 \label{fig:width-chi-piby2}
\end{figure}
%%%%%%%%%%%%%%%%%%%%%%%%%%%%%%%%%%%%%%%%%%%%%%%%%%%%%%%%%%%%%
%%%%%%%%%%%%%%%%%%%%%%%%%%%%%%%%%%%%%%%%%%%%%%%%%%%%%%%%%%%%%
\begin{figure*}[!htb]
\centering
\includegraphics[width=\linewidth]{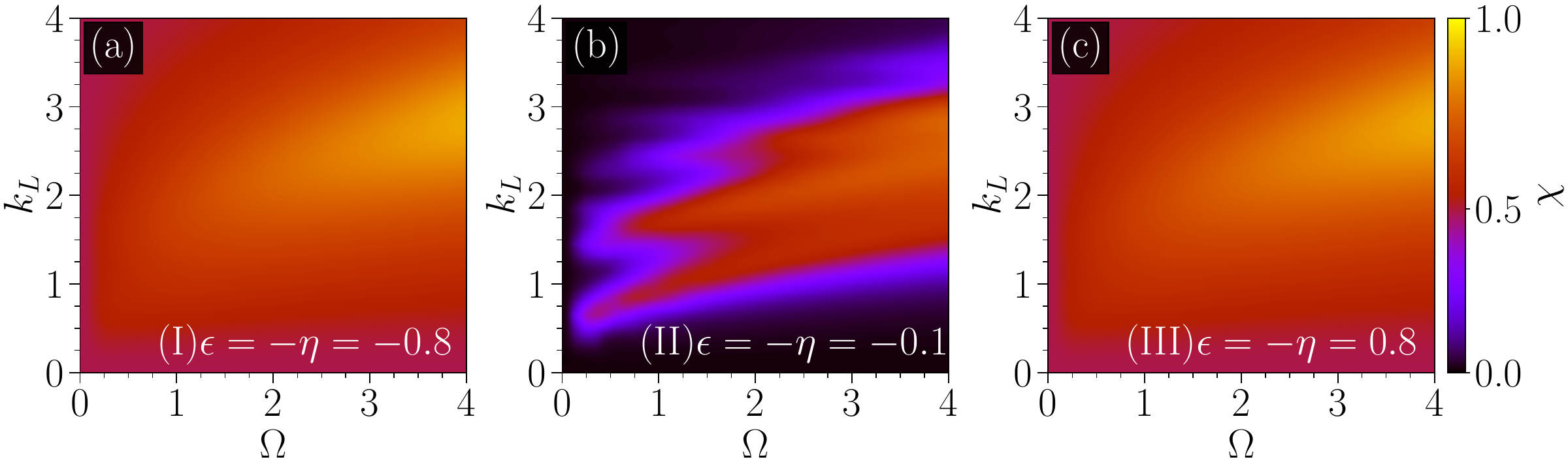}
\caption{Pseudo color representation of the form factor in the $\Omega-k_L$ plane in three distinct regions (I, II, and III) of interaction inhomogeneity parameter space as shown in the Fig.~\ref{fig:phase-ep-eta-pib2}: (a) $\epsilon = -\eta = 0.8$, region I, (b)$\epsilon = -\eta = -0.1$, region II, and (c) $\epsilon = -\eta = -0.8$, region III. Both regions I and III exhibit the localization for all ranges of $\Omega$ and $k_L$, while region II displays localization-delocalization transition by varying $k_L$ for $\Omega$ close to $1$.
}
\label{fig:phase-kL-Omega-piby2}
\end{figure*} 
%%%%%%%%%%%%%%%%%%%%%%%%%%%%%%%%%%%%%%%%%%%%%%%%%%%%%%%%%%%%%

Next, to get a comprehensive picture of the localized and delocalized phases due to the effect of the inhomogeneities, we show the pseudo-color representation of $\chi$ in the $\epsilon-\eta$ plane for $\Omega=1$ and $k_L=0.4$ in Fig.~\ref{fig:phase-ep-eta-pib2}. We noted that the delocalized region, where $\chi$ remains near zero, is sandwiched between two localized regions with $\chi \gtrsim 0.5$. This particular feature of the localization-delocalization transition with interaction inhomogeneities closely resembles the reentrant localization feature in previous studies using the Aubr\"{e}-Andre models~\cite{Roy:2021}. Note that as we consider $\theta_1=\theta_2=\pi/2$ the symmetry of the Hamiltonian remains intact, and so the localization-delocalization-localization appears to be symmetric with respect to $\epsilon\rightarrow-\epsilon, \eta\rightarrow -\eta$. However, this symmetry breaks as we destroy the symmetry of the Hamiltonian by considering $\theta_1\ne\theta_2$. In ~\ref{appen:B}, we provide one example for $\theta_1\ne\theta_2$ to show the asymmetric nature of the transition.

Further, we analyze the effect of spin-orbit coupling parameters on the localization-delocalization. In Fig.~\ref{fig:width-chi-piby2}(a), we show the variation of form factor $\chi$ with the interaction inhomogeneity $\epsilon(=-\eta)$ at three different values of $k_L = 0.0, 0.4$ and $0.6$. Conversely, Fig.~\ref{fig:width-chi-piby2}(b) illustrates the condensate width $w$ for the same values of $k_L$. In figure (a), for the case of $k_L = 0.4$, one can notice that the form factor increases between $0.4 \leq \chi \leq 0.6$ in between $\epsilon(=-\eta) \lesssim -0.5$ and $\epsilon(=-\eta) \gtrsim 0.5$, resembling the localized regions. On the other hand, the $\chi$ value remains almost constant ($\chi \sim 0$) in between $-0.5 \lesssim \epsilon(=-\eta) \lesssim 0.5$ depicting the delocalized region. Conversely, the condensate width $w$ complements the form factor $\chi$ by showing $w \lesssim 1$ in the localized regions, and in the delocalized region the width becomes $w \gg 1$. Another notable fact is that with the increment of $k_L$ from zero, the minimum value of $\chi$ in the delocalized region and the maximum value in the localized region get increased. For instance, in the delocalized region, the $\chi$ value remains in the order of $\sim \mathcal{O}(10^{-2})$, whereas the localized region is featured with the larger $\chi$ value greater than $ > 0.5$. Also, one may notice that the delocalized region (grey shaded region) shrinks as the $k_L$ value increases with the variation of $\epsilon (=-\eta)$. On the other hand, the condensate width $w$ (dashed line with open markers) nicely complements the form-factor behaviour by showing a larger value of $w \gtrsim 10.0$ in delocalized region and the smaller value $w \lesssim 1.0$ in localized regions.

In order to get more comprehensive picture, we construct the phase diagram in $k_L - \Omega$ plane by keeping $\epsilon$ and $\eta$ fixed at different regions, i.e., (I), (II) and (III) as shown in Fig.~\ref{fig:phase-ep-eta-pib2}. The phase diagram in Fig.~\ref{fig:phase-kL-Omega-piby2}(a) corresponds to the localized region at (I) ($\epsilon=-\eta = 0.8$) where the $\chi$ value lies within the range $0.49 \lesssim \chi \lesssim 0.88$, clearly demonstrating the localized feature of the condensate. On the other hand, when we decrease the interaction parameters towards delocalized region such as region (II) ($\epsilon = - \eta = -0.1$), the associated form-factor in Fig.~\ref{fig:phase-kL-Omega-piby2}(b) exhibits similar localization-delocalization phenomenon as shown previously in Fig.~\ref{fig:phase-omega-kL}(b). Next, Fig.~\ref{fig:phase-kL-Omega-piby2}(c) corresponds to the region (III) ($\epsilon = -\eta = -0.8$), which shows the localized condensate for all the range of coupling parameters.

So far, our investigations have mainly focused on numerical analysis through which we have characterized various localized and delocalized regions with variations in interactions as well as the coupling strength of the system. To understand localization-delocalization transitions, either by tuning the coupling parameters or changing the interaction strengths, we adopt the Gaussian variational approach in the following.
%So far, our investigations have been mainly focused on the numerical analysis through which we have characterized various localized and delocalized regions with the variation of interactions as well as the coupling strength of the system. To understand the reentrant localization, either tuning the coupling parameters or changing the interaction strengths, we adopt the Gaussian variational approach in the following.
%%%%%%%%%%%%%%%%%%%%%%%%%%%%%%%%%%%%%%%%%%%%%%%%%%%%%%%%%%%%%%%%%
\begin{figure}[!htp]
 \centering
 \includegraphics[width=0.8\linewidth]{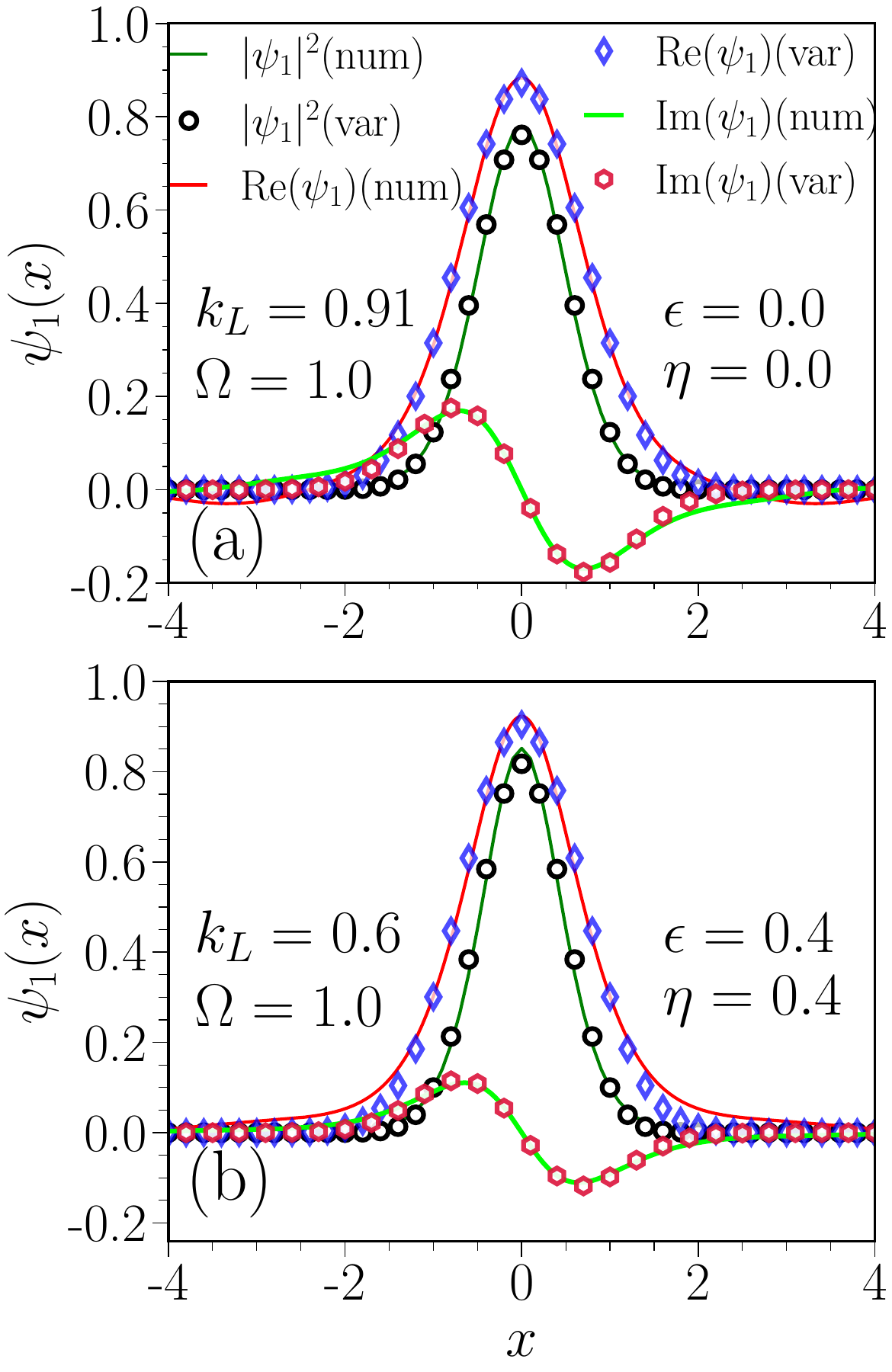}
\caption{Ground state density profile obtained using numerical (with lines) and variational scheme (with markers) for (a) $k_L = 0.91$, $\epsilon (=\eta) = 0$, and (b) $k_L = 0.6$, $\epsilon (=\eta) = 0.4$. The real and imaginary parts of the spin-up condensate wavefunction obtained from the numerical simulation and variational approach agree well.}
 \label{density-gamma-0p91-TD}
\end{figure}
%%%%%%%%%%%%%%%%%%%%%%%%%%%%%%%%%%%%%%%%%%%%%%%%%%%%%%%%%%%%%

%===================================================%
\subsection{Role of inhomogeneity in the localization-delocalization transition}
\label{sec:III_B}
In this section, using the variation approach, we present a plausible explanation for the observation of localization-delocalization obtained using the numerical simulation, as presented in Sec.~\ref{sec:C}. In what follows, we discuss the analytical model using the variational approach to disentangle the role of different localizing and delocalizing factors in spatially inhomogeneous SO coupled BECs.
%In this section, using the variation approach we present a plausible explanation for the observation of reentrant localization obtained using the numerical simulation as presented in Sec.~\ref{sec:C}. In what follows, we discuss the analytical model using the variational approach that has been utilized to disentangle the role of different localizing and delocalizing factors in spatially inhomogeneous SO coupled BECs. 

\subsubsection{Variational analysis}
\label{subsec:var-anal}

Using the time-dependent variational approach~\cite{cheng:2014}, the Lagrangian density of Eqs.~(\ref{eqn1(a)}) and (\ref{eqn1(b)}) can be written as,
\begin{align}
\mathcal{L} = & \sum_{j = 1}^2 \Bigg\{\frac{i}{2}\left(\psi_j^{\ast}\dot{\psi_j} - \psi_j\dot{\psi_j^{\ast}}\right) - (-1)^j\frac{i}{2}k_L \left[\psi_j^{\ast}\psi_j^{\prime} - \psi_j(\psi_j^{\ast})^{\prime}\right] \notag \\
& \left. - \frac{1}{2}\vert \psi_j^{\prime}\vert ^2 - \frac{1}{2}g(x)\vert \psi_j\vert ^4 - V(x)\vert \psi_j\vert ^2 \Bigg\}\right. \notag \\
& \left. - g_{12}(x)\vert \psi_1\vert ^2\vert \psi_2\vert ^2 - \Omega\left(\psi_1^{\ast}\psi_2 + \psi_1\psi_2^{\ast}\right)\right.
\label{eqn6}
\end{align}
where asterisk denotes the complex conjugate, prime represents the derivative ($d / dx$), and overdot denotes the time derivative ($d / dt$). Here, the notation $j = 1$ and $2$ refers to the spin-up and spin-down components, respectively.

%%%%%%%%%%%%%%%%%%%%%%%%%%%%%%%%%%%%%%%%%%%%%%%%
%\subsection{Time dependent variational approach}
To obtain the corresponding dynamical equations, we use the Gaussian ansatz with time-dependent variational parameters $N_j$, $w_j$, $\beta_j$, $x_{0_{j}}$ and $\phi_j$ of the form:
\begin{align}
 \begin{split}
 \psi_j(x,t) = & \frac{1}{\pi^{1/4}}\sqrt{\frac{N_j}{w_j}}\exp \Bigg[-\frac{(x-x_{0_j})^2}{2w_j^2}\Bigg] \\
 & \left. \times \exp \Bigg[(-1)^j i \beta_j(x-x_{0_j}) + i \phi_j\Bigg] \right.,
 \end{split}
 \label{eqn7}
\end{align}
%where $j = 1, 2$ corresponds to the aforementioned spin-up and spin-down components, respectively. 
where $N_j$ is the number of atoms, $w_j$ represents the width of the condensate centered at $x_{0_j}$, $\beta_j$ and $\phi_j$ are respectively the chirp and phase of the condensate with Gaussian profile. At this stage, it will be worthwhile to point out that the variational ansatz is more appropriate to comprehend the characteristics of the localized region of the condensate, not the behavior of the delocalized regions. To complement the delocalized state using the variational analysis, one needs to consider the superposition of more than one Gaussian wavefunction that will contain a large number of variational parameters and will come with added complexities. 
The effective Lagrangian with the Gaussian ansatz can be obtained by substituting Eq.~(\ref{eqn7}) into Eq.~(\ref{eqn6}) and integrating over space variable with the potential Eq.~(\ref{eqn3}) and the nonlinearity Eq.~(\ref{eqn5}). The detailed calculation for the dynamical equation of motion using the Lagrangian given in ~\ref{Appen:1}. 
\begin{widetext}
\begin{subequations}
\label{eqn8}
\begin{align}
	\begin{split}
	 k_L -\beta - \sqrt{\frac{N_{3-j}}{N_j}}\hspace{2mm}\frac{\partial L_{\Omega}}{\partial \beta_j} = 0
	\end{split}
	\label{eqn8a}
	\end{align}
\begin{align}
% \begin{split}
 & \frac{1}{2 w^{3}} +\frac{1}{2\sqrt{2\pi} w^{2}}\Bigg[\epsilon_{0}-\frac{\epsilon}{4} \sum_{l=1}^{2} V_{l}\left(k_{l}^{2} w^{2}+4\right) \exp(-\frac{k_{l}^{2} w^{2}}{8}) 
\times \cos \left(k_{l} x_{0}\right)\Bigg] -\frac{w}{2} \sum_{l=1}^{2}\left(V_{l} k_{l}^{2}\right) \cos \left(k_{l} x_{0}\right) \exp \left(-\frac{k_{l}^2 w^2}{4}\right) \notag \\ & 
+\frac{\eta_{0} }{2\sqrt{2\pi}w^4 }\left(w^{2}-4x_{0}^{2}\right) \exp \left(-\frac{2x_{0}^2}{w^{2}}\right) 
-\frac{\eta}{2 \sqrt{2\pi}w^{4}} \sum_{l=1}^{2} V_{l} \left(-4x_{0}^{2}+ w^{2} + \frac{k_{l}^{2} w^{4}}{4}\right) \exp (-\frac{k_{l}^{2} w^{2}}{8}) \notag \\ &
+ 2\Omega \exp \left( -\frac{x_0^2 + \beta^2w^4}{w^2}\right)\left(\beta^2w + \frac{x_0^2}{w^3}\right)=0
 \label{eqn8b}
\end{align}
%EOM with beta_j stationary condition
\begin{align}
 & \frac{\textcolor{black}{(-1)^{j-1}}N_{j} \epsilon}{2 \sqrt{2\pi} w} \sum_{l=1}^{2}\left(V_{l} k_{l}\right) \sin \left(k_{l} x_{0}\right) \exp \left(-\frac{k_{l}^{2} w^{2}}{8}\right) 
+ \sum_{l=1}^{2}\left(V_{l} k_{l}\right) \sin \left(k_{l} x_{0}\right) \exp \left(-\frac{k_{l}^{2} w^{2}}{4}\right) -\frac{\sqrt{2} \eta_{0} N_{3-j}x_{0}}{\sqrt{\pi}w^{3}} \notag \\ &
\times \exp \left(-\frac{2x_{0}^2}{w^{2}}\right) -\frac{\textcolor{black}{(-1)^{2j-1}}\eta N_{3-j}}{2^{3/2}\sqrt{\pi}w^{3}} \exp \left(-\frac{2x_{0}^{2}}{w^{2}}\right) 
\sum_{l=1}^{2}(-1)^{j}4V_{l} x_{0} \exp \left(-\frac{k_{l}^{2} w^{2} }{8}\right) + \sqrt{\frac{N_{3-j}}{N_j}} (-1)^{j-1} \frac{\partial L_{\Omega}}{\partial x_{0_{j}}} = 0
\label{eqn8c}
\end{align}
\end{subequations}
\end{widetext}

First, we focus on analyzing the stationary state upon solving the coupled equations (Eq.~(\ref{eqn1(a)}) and Eq.~(\ref{eqn1(b)}) ) considering equal atoms in both components (i.e., $N_1 = N_2$). In order to obtain the stationary conditions, we need to make the time derivative of Eqs.(\ref{eqn:A2b}–\ref{eqn:A2d}) is equal to zero. As a result, we obtain $\beta_1 = \beta_2 = \beta$, $w_1 = w_2 = w$, and $x_{0_{1}} = -x_{0_{2}} = x_0$ as the initial condition. Using these initial conditions, the equations of motion associated to $x_0$, $w$ and $\beta$ have the form,

%%%%%%%%%%%%%%%%%%%%%%%%%%%%%%%%%%%%%%%%%%%%%%%%%%%%%%%%%%%%%%%%%%%%%
\begin{figure}
 \centering
 \includegraphics[width=\linewidth]{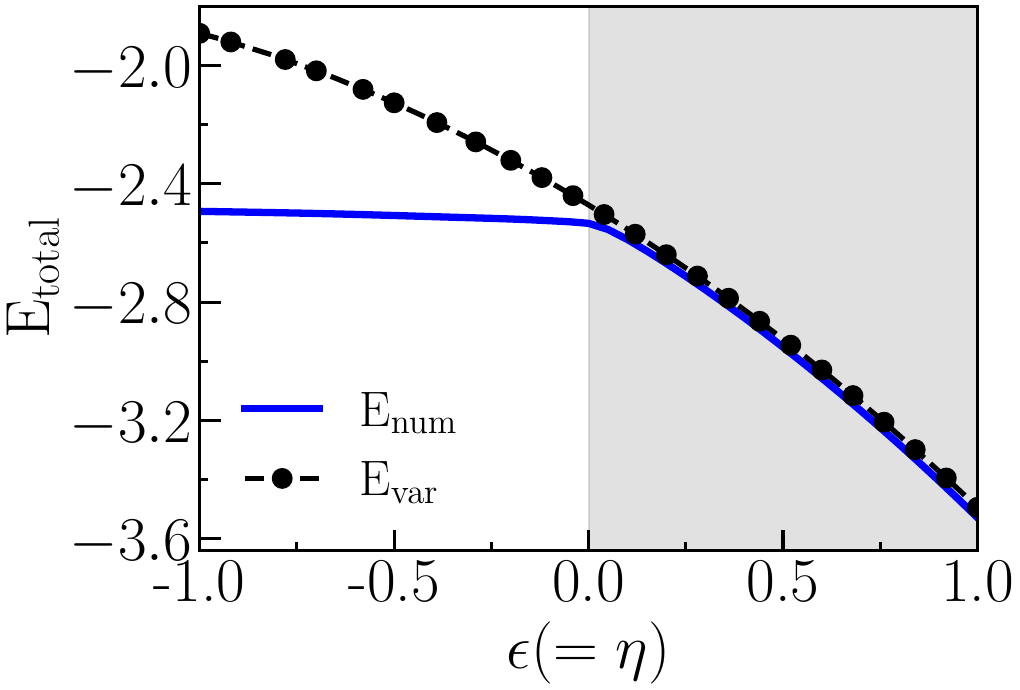}
 \caption{Variation of the ground state energy as a function of the inhomogeneity parameter ($\epsilon = \eta$). $E_{\rm num}$ (solid-blue line) and $E_{\rm var}$ (dashed line with dots) represent the total energy obtained from numerical simulation and analytical variational approach, respectively. Here $k_L = 0.6$ and $\Omega = 1$. There is good agreement between the numerical and variational energy in the localized region ($\epsilon=n>0$) where energy increases towards more negative, indicating the formation of stronger localization with an increase in the inhomogeneity. In the delocalized region, $E_{\rm num} \sim -2.5$ for negative inhomogeneity parameter. }
 \label{fig:Energy-compare-wo-phase}
 \end{figure}
%%%%%%%%%%%%%%%%%%%%%%%%%%%%%%%%%%%%%%%%%%%%%%%%%%%%%%%%%%%%%%%
In Fig.~\ref{density-gamma-0p91-TD} we show a comparison between the density profile of spin-up component obtained using {GPEs} (with solid line) and that using the variational approach [Eq.~(\ref{eqn8a})–(\ref{eqn8c})
] shown with open markers considering $k_L = 0.91$, $\epsilon = \eta = 0$, and $\Omega = 1.0$ [see Fig.~\ref{density-gamma-0p91-TD}(a)], and $\epsilon = \eta = 0.4$, $k_L = 0.6$ and $\Omega = 1.0$ [see Fig.~\ref{density-gamma-0p91-TD}(b)]. At stationary state, $\vert \psi_{1}\vert^2 = \vert \psi_{2}\vert^2$, $\mathrm{Im} \, \psi_{1} = \mathrm{Im} \, \psi_{2}$, and $\mathrm{Re} \, \psi_{1} = -\mathrm{Re} \, \psi_{2}$. Noting this, hereafter, we consider the spin-up component for the discussion. The density profile calculated using the numerical {GPEs} and variational approach agree well in the localized state. For the non-interacting case ($\epsilon = \eta = 0$), the variational approach (Eq.~(\ref{eqn8b})) reveals that the kinetic energy (first term) is responsible for the delocalization. However, the Rabi-coupling and the optical lattice potential contribute to localization of the condensate, which is in line with the observations made earlier (refer to the discussion in Sec.~\ref{sec:1A}). Further, to highlight the effect of inhomogeneities on the localization in Fig.~\ref{density-gamma-0p91-TD}(b), we illustrate the density profile for $\epsilon = \eta = 0.4$, $k_L = 0.6$ and $\Omega = 1.0$. The density profile obtained using the variational approach agrees well with those obtained using {GPEs} with slightly higher amplitude than those for non-interacting cases, indicating a stronger localization with repulsive nonlinearity.

In Fig.~\ref{fig:Energy-compare-wo-phase} we show the variation of total energy ($E_{\rm total}$) with interaction inhomogeneity as ($\epsilon=\eta$). For attractive interaction ($\epsilon =\eta \leq 0$), the total energy calculated using numerics remains constant at $E_{\rm num} = -2.5$. On the other hand, for repulsive interactions ($\epsilon(=\eta) \geq 0$), the $E_{\rm num}$ decreases monotonically upon increasing interaction strength, indicating that the localized condensate minimizes the energy. For negative inhomogeneous interaction, the total energy $E_{\rm var}$ calculated using the variational approach fails to agree with those obtained using the numerical simulation. The discrepancy between the variational and numerical simulation results for the delocalized region can be attributed to the assumption of Gaussian-like ansatz for the wave function in the variational analysis. One needs to consider the superposition of a large number of Gaussian states to emulate the behaviour of a delocalized state, which will make the variational calculation quite complex.

%\subsubsection{Role of inhomogeneity in the reentrant of localization}

The localization appears to arise from the interplay between interactions and the bichromatic optical lattice in a collisionally inhomogeneous environment. To account for this, we consider a phase difference of $\theta=\pi/2$ between the interaction and the potential, as given in Eqs.~(\ref{eqn6a}) and (\ref{eqn6b}), respectively, which we further substitute into the Lagrangian (Eq.~(\ref{eqn6})). Using the Euler-Lagrange formalism, we calculate the equations of motion with respect to the variational parameters. The details of the Lagrangian are provided in Appendix~\ref{appen:B}.
%The localization appears as a consequence of the interplay between interactions and the bichromatic optical lattice in a collisionally inhomogeneous environment. To take account of these, the phase $\theta=\pi/2$ between the interaction and the potential is considered to be of the form as given in the Eq.~(\ref{eqn6a}) and Eq.~(\ref{eqn6b}), respectively which we further substitute in the Lagrangian (\ref{eqn6}). Using the Euler-Lagrange formalism the equations of motion with respect to the variational parameters are calculated. The detailed form of the Lagrangian is supplied in Appendix~\ref{appen:B}.

\subsubsection{Effective Potential for localization-delocalization of the condensate}
\label{subsec:eff_pot}

To understand the role of inhomogeneity in the localization of the condensate, we resort to derive the dynamical equation of the condensate moving under the influence of effective potential mainly generated due to the spatial inhomogeneity~\cite{Cheng:2010, Cheng:2011}. 

The effective potential can be obtained using Eq.~(\ref{eqn:A2c}) and Eq.~(\ref{eqn:A2e}) (see Appendix \ref{Appen:1} and \ref{appen:B} for details), which yields:
%%%%%%%%%%%%%%%%%%%%%%%%%%%%%
\begin{align}
 \frac{\partial^2 x_{0_j}}{\partial t^2} = - \dot{\beta}_j = -\frac{\partial V^{\mbox{eff}}_j}{\partial x_{0_{j}}} \label{eqn22}
% \frac{\partial^2 x_{0_j}}{\partial t^2} = \label{eqn23}
\end{align}
%%%%%%%%%%%%%%%%%%%%%%%%%%%%%
%%%%%%%%%%%%%%%%%%%%%%%%%%%%%
\begin{figure}
 \includegraphics[width=\linewidth]{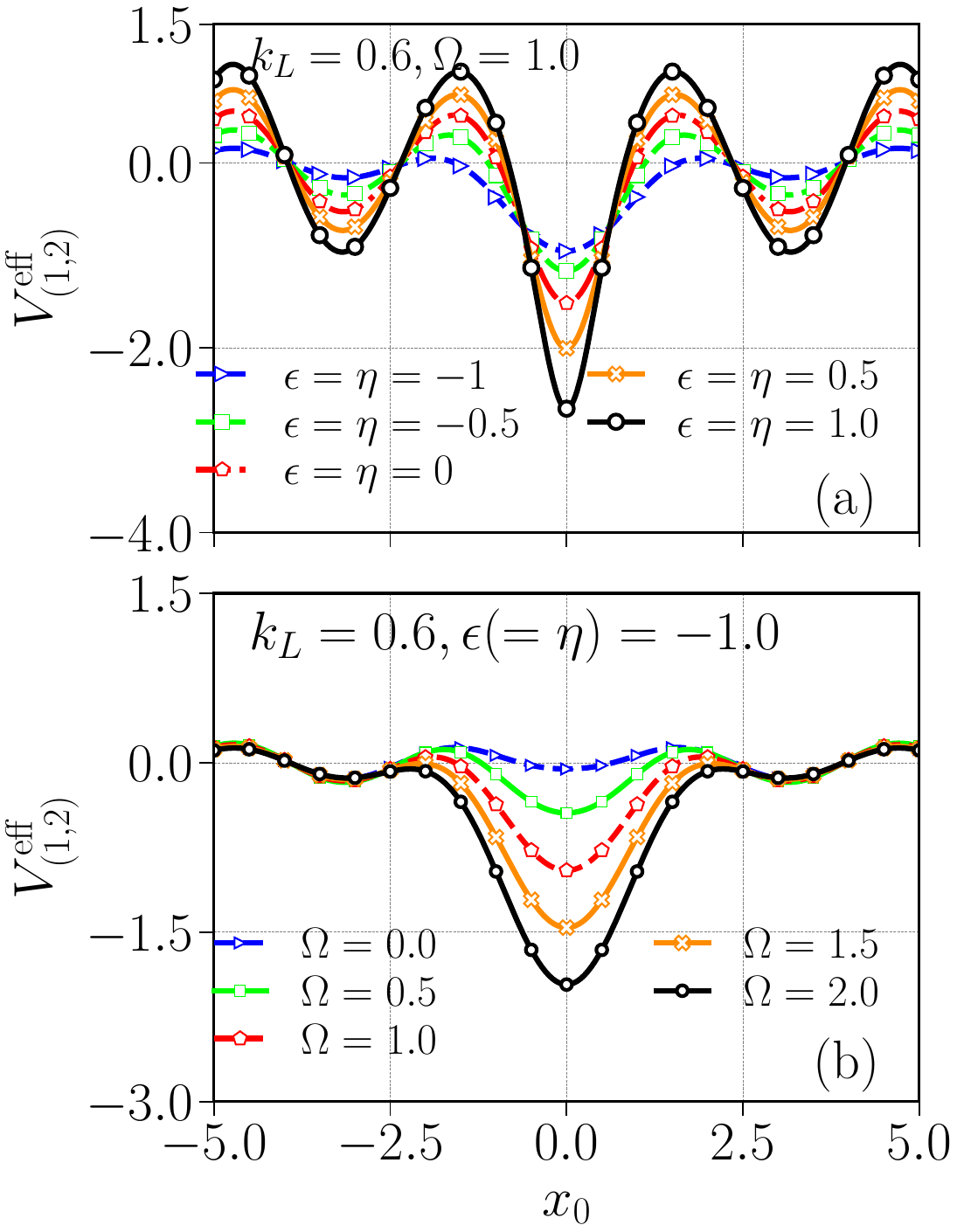}
 \caption{Effective potential ($V^{\rm eff}_{(1,2)}$) profile for the condensate obtained using variational approach [Eq.~\ref{eqn:A3a}]: (a) for different inhomogeneous interaction parameter keeping $\Omega=1$ and $k_L=0.6$ and (b) for different $\Omega$ keeping $k_L=0.6$ and inhomogeneity interaction as $\epsilon(=\eta)=-1.0$. The depth of the effective potential at the central point ($x_0\sim0$) increases upon increasing the repulsive interaction compared to those for $\epsilon=\eta=0$ indicating the localization for the condensate for repulsive interaction inhomogeneity. However, the trend is opposite for an increase in the attractive interaction, implying the delocalization for them. Upon an increase of $\Omega$ in (b), the central depth increases, suggesting the localization of condensate for this increment. }
 \label{fig:eff_potential}
\end{figure}
%%%%%%%%%%%%%%%%%%%%%%%%%%%%%%%%%%%%%%
%%%%%%%%%%%%%%%%%%%%%%%%%%%%%%%%%%%%%%
\begin{figure}
 \centering
 \includegraphics[width=\linewidth]{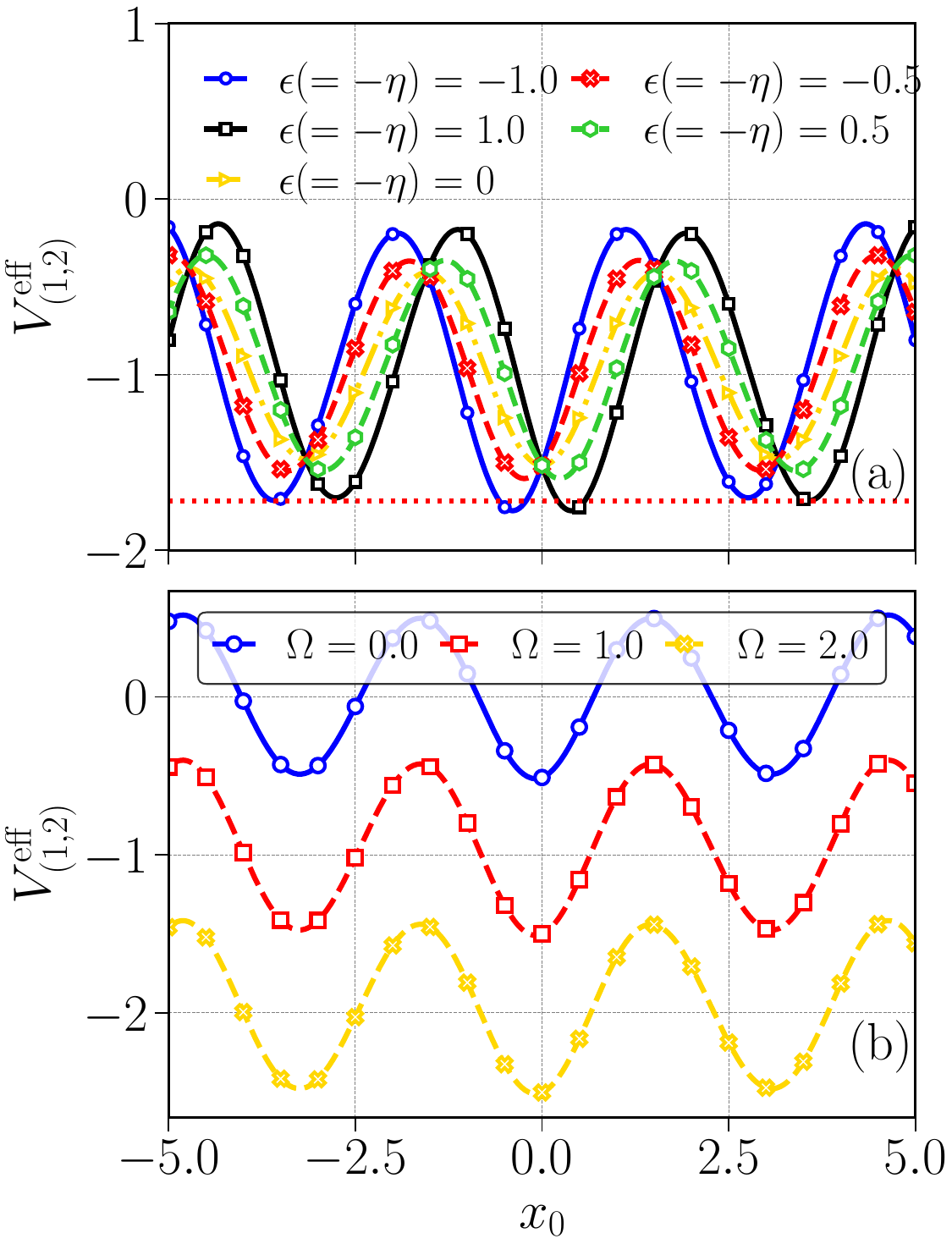}
\caption{(a) Variation of effective potential ($V^{eff}_{(1,2)}$) as obtained in Eq.~\ref{eqn:B1b} of spin-components with $x_0$ for different inhomogeneous interaction parameters $\epsilon(=-\eta)$ when the phase difference between the spatial inhomogeneity and potential considered is $\pi/2$. For $\epsilon=\eta=0$, the potential is peaked at $x_0=0$, which is the delocalized state. However, for $\lvert \epsilon(-\eta) \rvert=0.5$ and $1$, the central minima depth increases compared to those for zero inhomogeneity, indicating the localization state. (b) Variation of effective potential with $x_0$ for different $\Omega$ with $k_L = 0.6$, $\epsilon=\eta = 0$. As the value of $\Omega$ increases, the depth of the central potential also increases, implying that the Rabi coupling contributes towards the localization of the condensate.
%(a) Variation of effective potential ($V^{eff}_{(1,2)}$) as obtained in Eq.~\ref{eqn:B1b} of spin-components with $x_0$ for different inhomogeneous interaction parameters $\epsilon(=-\eta)$ when the phase difference between the spacial inhomogeneity and potential considered is $\pi/2$. For $\epsilon=\eta=0$ the potential is peaked at $x_0=0$ which is the delocalized state, however, for $|\epsilon(-\eta)|=0.5$ and $1$ the central minima depth increases compared to those for zero inhomogeneity indicating the localization state. (b) Variation of effective potential with $x_0$ for different $\Omega$ with $k_L = 0.6, \epsilon=\eta = 0$. Increase in $\Omega$ results in increasing the depth of the central potential implying that the Rabi-coupling contributes towards localization of the condensate.
} 
\label{fig:eff_potential_theta}
\end{figure}
%%%%%%%%%%%%%%%%%%%%%%%%%%%%%%%%%%%%%%

In Fig.~\ref{fig:eff_potential}(a), we show the profile of $V_{\rm eff}$ corresponding to the spin-up component upon varying the inhomogeneous interactions while keeping the intra and intercomponent interaction same for $k_L=0.6$ and $\Omega=1$. We find that the effective potential attains a global minimum at $x_0 = 0$, and the depth of the global minimum increases upon an increase in the spatial repulsive inhomogeneous interaction. For instance, at $\epsilon =\eta = 1$ (open circled black solid line), a larger depth of effective potential ($V_{\rm eff} \sim 2.5$) indicates that the condensate gets more localized as the repulsive interaction increased. However, for attractive interaction($\epsilon$ or $\eta < 0$), the depth of minima at $x_0\sim 0$ becomes of the order of other minima of the effective potential, indicating the delocalized state of the condensate. This observation aligns with the earlier numerical results obtained with the one component {BECs}~\cite{Cheng:2011}. Additionally, this result closely matches the observed transition from localized to delocalized states when interaction parameters are varied from positive to negative (see Fig.~\ref{fig:xhi-width-compare}). 
%In Fig.~\ref{fig:eff_potential}(a) we show the profile of $V_{\rm eff}$ corresponding to the spin-up component upon varying the inhomogeneous interactions while keeping the intra and intercomponent interaction same for $k_L=0.6$ and $\Omega=1$. We find that the effective potential attains a global minimum at $x_0 = 0$ and the depth of the global minimum increases upon increase in the spatial repulsive inhomogeneous interaction. For instance at $\eta=\epsilon=1$ (open circled black solid line), larger depth of effective potential ($V_{\rm eff}\sim 2.5$) indicates that the condensate gets more localized as the repulsive interaction increased. However, for attractive interaction($\epsilon$ or $\eta < 0$) the depth of minima at $x\sim 0$ becomes of the order of other minima of the effective potential indicating of delocalized state of the condensate. This particular observation is in line with the earlier numerical results obtained with the one component BECs~\cite{Cheng:2011}. In addition this feature agrees quite well with the numerical observation of the localization to delocalization transition upon varying the interaction parameters from positive to negative (See Fig.~\ref{fig:xhi-width-compare}). 

%%%%%%%%%%%%%%%%%%%%%%%%%%%%%%%%%%%%%%%%%%%%%%%%%
\begin{figure*}[!htb]
 \centering
\includegraphics[width=0.9\linewidth]{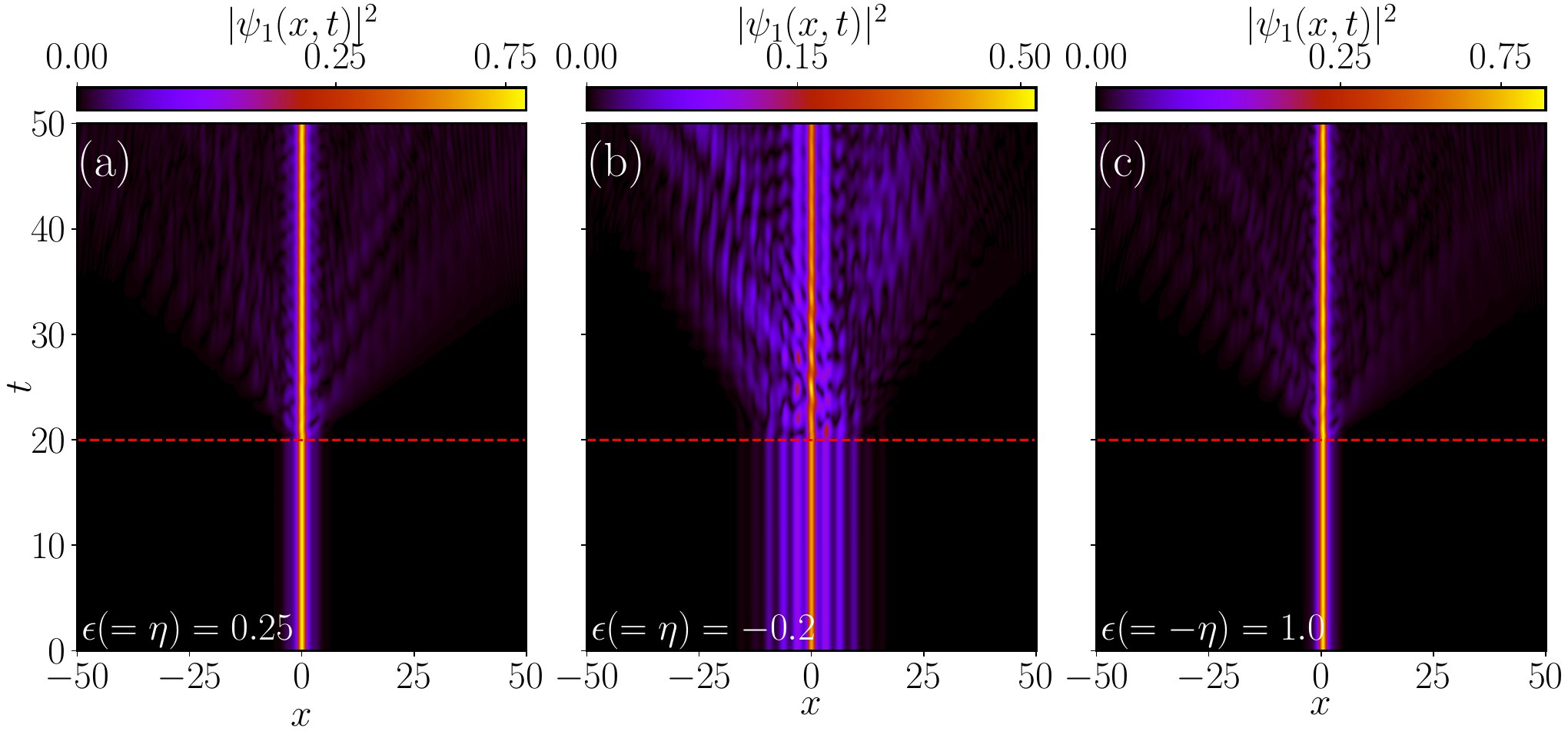}
\caption{Pseudo color representation of the velocity-induced dynamics of the spin-up component density for localized (a) and delocalized (b) state of the case when the inhomogeneity has the same phase as the potential. (c) shows the localized state for the case when the inhomogeneous interaction and potential have $\pi/2$ phase difference. The other parameters are $k_L = 0.6$ and $\Omega = 1$. Both the localized states (a) and (c) do not show any change in shape and size upon perturbing with the velocity, while the delocalized state (b) shows the spread in shape with time. }
\label{fig:dynamics_vel-0p2}
\end{figure*}
%%%%%%%%%%%%%%%%%%%%%%%%%%%%%
In the similar line as discussed above, to understand the role of Rabi coupling in attaining the localization, we show the effective potential in Fig.~\ref{fig:eff_potential}(b) as $\Omega$ is increased for the condensate with $\epsilon=\eta=-1$ and $k_L=0.6$. For $\Omega=0$, the minima of the effective potential at $x_0=0$ is of the same width as of minima at other $x_0$, indicating the delocalized state of the condensate. As the Rabi-coupling increased from zero, the depth of effective potential at $x_0 =0$ increases and becomes significantly higher compared to the minima at other $x_0$ for $\Omega\gtrsim 0.5$. This variation of the effective potential with increasing $\Omega$ indicates that the role of Rabi coupling is to localize the condensate in the presence of inhomogeneous interaction where it was initially delocalized in line with the numerical observation made in the Sec.~\ref{sec:1A}. 

Next, we turn our focus on analyzing the role of spatial inhomogeneity on the localization-delocalization that arises when the inhomogenous nonlinear interaction carries the $\pi/2$ phase difference in spatial dependence as those for the trapping quasiperiodic potential as described in Sec.~\ref{sec:C}. The form of the effective potential is provided in Appendix~\ref{appen:B}.

Interestingly, the spatial variation of the effective potential in Fig.~\ref{fig:eff_potential_theta}(a) vividly illustrates the shifting of the minima with the increment of nonlinear interactions $\epsilon (= -\eta)$ as a result of the captivating interplay between the potential and spatial dependence of the interaction. For instance, as we move from $\epsilon= \eta = 0$, the local minima at $x_0 = 0$ shifts either in the direction of positive or negative $x_0$, as depicted for $\epsilon = -\eta = 0.2$ (green dashed line) or $\epsilon = -\eta = - 0.2$ (purple dashed line), respectively. Notably, the displacement of the minima occurs symmetrically on both sides of the center. For example, at $\epsilon (=-\eta) = 1.0$, the minima are displaced around $x_0 = 0.3$, whereas at $\epsilon (=-\eta) = -1.0$, the minima are found to be at $x_0 = -0.3$. Also, the presence of disorder in the potential leads to a slightly larger depth of the potential well located at $x_0 = \pm 0.3$ compared to the other wells for the aforementioned interactions, which is clearly distinguishable by the red dotted line in the figure. These particular features highlight intriguing competition between the potential and nonlinearity towards the localization-delocalization behavior. Additionally, the effect of Rabi-coupling towards localization is clearly evident when studying the magnitude of $V_{\rm eff}$ in Fig.~\ref{fig:eff_potential_theta}(b). 
%%%%%%%%%%%%%%%%%%%%%%%%%%%%%%%%%%%
%\section{results}
%\subsection{Stationary localized state}
%%%%%%%%%%%%%%%%%%%%%%%%%%%%%%%%%%%%

\subsection{Effect of spatial inhomogeneity on the dynamics of localized and delocalized states}
\label{Sec:D}
So far, we have analyzed the effect of the SO and Rabi coupling along with the interaction inhomogeneity on the localization of the condensate and studied their role using the variational approach. In this section, we present the dynamics of those localized and delocalized states by perturbing the ground state through velocity or performing a quench on the strength of the secondary potential. 
%\sout{using two means: (i) by quenching the potential and (ii) by considering atomic imbalance between two components as an initial state.}
\subsubsection{Velocity perturbed Dynamics }
In this section, we present the dynamics that ensued in the condensate as a result of the finite equal and opposite velocity imparted to the condensate in the localized or delocalized state in the presence of the interaction inhomogeneities. After imparting the velocity, we analyze the condensate dynamics utilizing the real-time propagation scheme.

%%%%%%%%%%%%%%%%%%%%%%%%%%%%%%%%%%%%%%%%%%%%%%%%%%%%%%%%%%%%%
In Fig.~\ref{fig:dynamics_vel-0p2}, we show the temporal evolution of spin-up component condensate density for various interaction inhomogeneities. In panel (a), we observe the initial localized state obtained when the phase between spatial inhomogeneity and potential is zero ($\theta = 0$), with parameters $\epsilon = \eta = 0.2$, $k_L = 0.6$, and $\Omega = 1.0$. After applying a velocity perturbation at $t = 20$, we observe a slight change in the condensate. Only a small fraction generates ripples in space, while the maximum density remains almost unchanged compared to the ground state density. A similar behaviour is observed for the localized condensate when the phase shift is $\theta = \pi/2, \epsilon(=-\eta) = 1.0$ (see panel (c)). In panel (b), the delocalized condensate density demonstrates expansion in space over time after the velocity perturbation at $t = 20$ (after the red dashed line). In summary, it can be said that localized condensates [see Fig.~\ref{fig:dynamics_vel-0p2}(a) and Fig.~\ref{fig:dynamics_vel-0p2}(c)] remain stable against small perturbations, while delocalized condensates [see Fig.~\ref{fig:dynamics_vel-0p2}(b)] exhibit instability under similar perturbations~\cite{Cheng:2011}.
%In Fig.~\ref{fig:dynamics_vel-0p2}, we show the temporal evolution of spin-up component condensate density for various interaction inhomogeneities. In panel (a), we observe the initial localized state obtained when the phase between spatial inhomogeneity and potential is zero ($\theta = 0$), with parameters $\epsilon = \eta = 0.2$, $k_L = 0.6$, and $\Omega = 1.0$. Subsequently, following the velocity perturbation applied at $t = 20$, we observe a slight change in the condensate, with only a minimal fraction generating ripples in space, while the maximum density remains nearly constant compared to the ground state density. A similar behavior is observed for the localized condensate, when the phase shift is $\theta = \pi/2, \epsilon(=-\eta) = 1.0$ (see panel (c)). In panel (b), the delocalized condensate density demonstrates expansion in space over time after the velocity perturbation at $t = 20$(after red dashed line). In summary, It can be said that the localized condensates (in Fig.~\ref{fig:dynamics_vel-0p2}(a) and in Fig.~\ref{fig:dynamics_vel-0p2}(c)) remain stable against small perturbations, whereas the delocalized condensates (in Fig.~\ref{fig:dynamics_vel-0p2}(b)) exhibit instability under similar perturbations~\cite{Cheng:2011}.

\subsubsection{Quench induced dynamics}

%%%%%%%%%%%%%%%%%%%%%%%%%%%%%%%%%%%%%%%%%%%%%%%%%%%%%%%%%%%%%%%%
\begin{figure}[!htb]
 \centering
 \includegraphics[width=\linewidth]{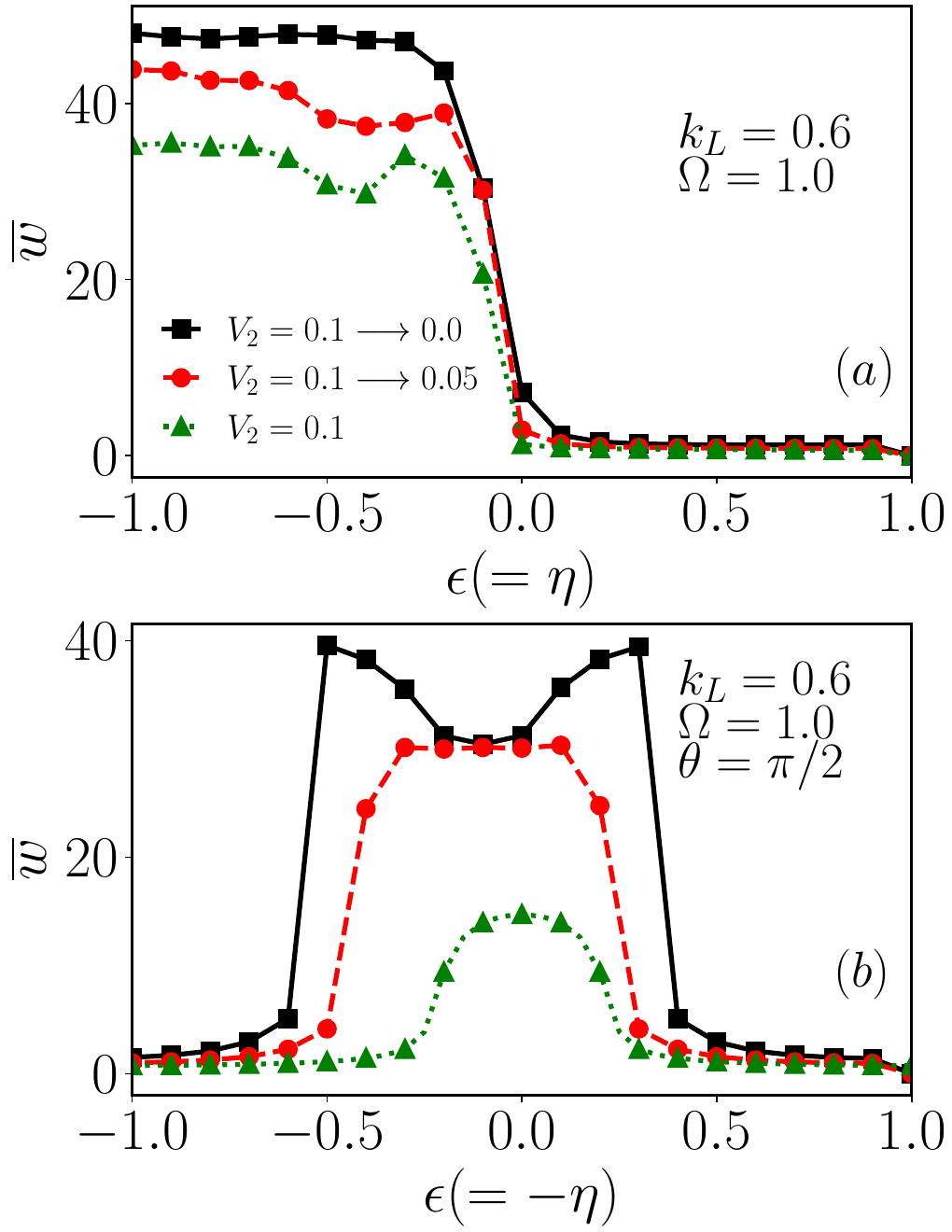}
 \caption{Time averaged mean-width ($\overline{w}$) of the condensate as a function of the inhomogeneity $(\epsilon=-\eta)$ for different quenching of the secondary potential ($V_2$) from initial value $0.1$ to the final value 0.0 (black rectangle) and 0.05 (red dots). Green triangle represents the case when $V_2=0.1$. The other parameters are $k_L = 0.6$, and $\Omega = 1.0$. For the case (a) when the inhomogeneity and potential have the same phase and (b) when the phase between the potential and nonlinear inhomogeneity is $\pi/2$.}
 
% Variation of time averaged condensate width ($\overline{w}$) for three different cases: without quenching (green triangle marked dotted line), quenching the quasiperiodicity of the trapping potential to lower value, i.e., $V_2 = 0.1 \rightarrow 0.05$ (red circle marked dashed line), and quenching the quasiperiodicity to zero, i.e. $V_2 = 0.1 \rightarrow 0.0$ (black square marked dashed line) in case of (a) without phase shift ($\theta_l = 0$), (b) with phase shift ($\theta_l = \pi/2k_l$). In both cases, it has been observed that in localized regions the condensate width remain approximately same after quenching the quasi periodicity in compare to the width estimated using ground state density, while in delocalized regions the width of the condensate increases significantly. Here, the spin-orbit and Rabi coupling parameters are considered to be $k_L = 0.6$, and $\Omega = 1.0$. }
 \label{fig:time-avg-evolution-post-trap-release}
\end{figure}
%%%%%%%%%%%%%%%%%%%%%%%%%%%%%%%%%%%%%%%%
Apart from the analysis of the localized and delocalized phases depending upon the ground state density with various coupling and interactions, now in this section, we explore the dynamics of the different phases of the condensate upon quenching of trap potentials~\cite{Ravisankar:2020, Ravishankar:2021, Gangwar:2024}. Firstly, we analyze the dynamics generated by applying the quench of the trap potential from initial to different final strength. In Fig.~\ref{fig:time-avg-evolution-post-trap-release}, we show the temporal average of the condensate width $\overline{w} = 1/T\int_0^T w(t) dt$, for quenching of the potential from initial ($V_2=0.1$) to lower trapping strength. In Fig.~\ref{fig:time-avg-evolution-post-trap-release}(a) we show the variation of mean width with $\epsilon(=\eta)$ as the dynamics are generated through two types of quenching: (i) by quenching the secondary lattice strength $V_2 = 0.1 \rightarrow 0.0$ (black solid line with square markers) and (ii) by quenching $V_2 = 0.1 \rightarrow 0.05$ (red dashed line with circular markers). We also show the $\overline{w}$ (green dotted line with triangular markers) corresponding to the ground state condensate. We find that for repulsive interaction $\epsilon(=\eta) > 0$, the $\overline{w}$ is of the same order compared to the $\overline{w}$ obtained without quenching. This particular feature implies that the condensate stability in the localized region is primarily caused by the repulsive inhomogeneity present in the system~\cite{Cheng:2011}.
%, irrespective of quasiperiodic (red dashed line) or periodic quenching (black solid line). 
On the other hand, with attractive interaction ($\epsilon(=\eta) < 0$), the $\overline{w}$ tend to increase towards a larger value to the order of $\overline{w} \sim \mathcal{O}(10^1)$. Moreover, in the case of periodic quenching (when final $V_2=0$), the mean width is higher compared to the case of quasiperiodic quenching (when $V_2=0.05$), which indicates that even a low but finite disorder is good enough to resist the expansion. {Note that as we use the equal interspecies interaction the density of the spin-up and spin-down component remains equal. Therefore, for the sake of brevity we have only showed the dynamical evolution of the spin-up component. The other component just follows up the spin-up component dynamics. }

%which leads to the slightly smaller value of $\overline{w}$ in compare to the quenching of $V_2$ from finite to zero. Further increment of the disorder strength also results in suppressing the condensate expansion with time. 

In Fig.~\ref{fig:time-avg-evolution-post-trap-release}(b), we present the mean condensate width with the variation of $\epsilon(=-\eta)$ in the presence of the phase shift $\theta = \pi/2$. Similar to $\theta = 0$ here also, we observe that in the localized region, $\overline{w}$ remains almost constant at the same order, $\mathcal{O}(10^{-1})$, irrespective of the strength of quenching of the trap. However in the delocalized region, the $\overline{w}$ follows similar behavior as observed for $\theta=0$ (See Fig.~\ref{fig:time-avg-evolution-post-trap-release}(a) ). In general, we find that the periodic quenching (final $V_2=0$) results in attaining the higher $\overline{w}$ compared to those for quasiperiodic quenching (final $V_2=0.05$).
%the value is lesser than the former case, although they become the same near to $\epsilon (= -\eta) \approx 0$. 
We also notice a slight shift of the delocalized regions towards the left side of $\epsilon (=-\eta) < 0$ for the delocalized regions compare to the case of without quenching (dotted line with green triangles). This feature can be attributed to the comparatively weak localization of the condensate with phase shift $\pi/2$ compared to the localization without phase shift. We also find the manifestation of the localization-delocalization transition with a change in inhomogeneity in the dynamics of the condensate captured through the variation of the mean condensate width in Fig.~\ref{fig:time-avg-evolution-post-trap-release}(b) in comparison to the results obtained using ground state in Fig.~\ref{fig:width-chi-piby2}(b).

%For the second case, we obtain the localized or delocalized ground state and quench the trap. In Fig. \ref{fig:time-evolution-post-trap-release} we show the variation of the $x_{rms}$ for three different sets of spatial inhomogeneity. (i) $\epsilon=\eta=-0.5$, (ii)$\epsilon=\eta=0$ and (iii) $\epsilon=\eta=0.5$. 

%%%%%%%%%%%%%%%%%%%%%%%%%%%%%%%%%%%%%%%%%%%%%%%%%%%%%%%%%%%%%
%\begin{figure*}[!htb]
% \centering
% \includegraphics[width=\linewidth]{rms_compare.pdf}
% \caption{Time evolution of the rms size of an expanding BEC released from harmonic trap into quasiperiodic (green solid line), periodic lattice (red solid) at $t = 70$~(red arrow marked) for ~(a) $\epsilon = \eta = -0.5, k_L = 0.6, \Omega = 1.0$,~(b) $\epsilon = \eta = 0.0, k_L = 0.91, \Omega = 1.0$,~(c) $\epsilon = \eta = 0.5, k_L = 0.6, \Omega = 1.0$. For comparison, the rms size of the free expansion BEC (blue dashed line) also plotted in each cases. }
% \label{fig:time-evolution-post-trap-release}
%\end{figure*}
%%%%%%%%%%%%%%%%%%%%%%%%%%%%%%%%%%%%%%%%%%%%%%%%%%%%%%%%%%%%%%%%%

\section{Conclusion}
\label{sec:IV}
We have numerically investigated the effect of interaction inhomogeneity on the localization-delocalization transition of the SO coupled binary quasi-one dimensional {BECs} trapped in the quasiperiodic potentials. By assuming the form of the spatial modulation of the interaction inhomogeneity same as that of the quasiperiodic potential we have shown that the condensed undergoes a localization to delocalization transition as the inhomogeneous interaction strength varies from attractive to repulsive. However, when a phase difference of $\pi/2$ is introduced between the inhomogeneity and the lattice potential, we have shown a remarkable localization-delocalization-localization transition as a function of the inhomogeneity strength.  Furthermore, we have also analyzed the effect of the SO and the Rabi coupling on the localization-delocalization transition for different strength of the inhomogeneity. For weak inhomogeneity strength, the increase in the SO coupling strength leads to localization-delocalization similar to the one observed for non-interacting homogeneous SO coupled {BECs}~\cite{Li:2016}.
%The delocalized and localized regions are characterized using the integral form factor and numerical width of the condensate. 
%The spatial dependence of the inhomogeneities is considered to be similar to that of the quasiperiodic potential with zero and $\pi/2$ phase difference. For the case when the potential and nonlinear inhomogeneity carries the same phase, the condensate shows the tendency of getting more and more localized upon increasing the inhomogeneity strength towards positive (repulsive) interaction and showing the delocalization tendency upon increasing the inhomogeneity strength towards negative (attractive interaction). 
%Further, we have also analyzed the effect of SO and Rabi coupling on the localization-delocalization transition for different strength of the inhomogeneity \sout{strength}. For weak inhomogeneity strength $(|\epsilon|\lesssim 0.1$) the increase in SO strength leads to reentrant  localization same as observed for non-interacting homogeneous SO coupled BECs~\cite{Li:2016}. 

%\tm{I have corrected the previous paragraph and the last paragraph. Please read the following paragraph carefully. it is not clear to me and it has several mistakes.}
Furthermore, we have utilized the variational approach to understand the competing nature of the quasiperiodic disorder due to the inhomogeneity and the potential responsible for the localization-delocalization-localization, which is revealed clearly in the oscillatory nature of the depth of the effective potential as a function of the inhomogeneity strength. We have also demonstrated the manifestation of the localization-delocalization in the quench dynamics of the condensate. Our present analysis provides an alternate disorder parameter other than the external potential to attain this behavior, which may motivate future experiments in {BECs}. 
%We have utilized the variational approach to understand the competing nature of the quasiperiodic disorder due to the inhomogeneity and the potential responsible for the renentrant localization which is clearly revealed in the oscillatory nature of the depth of the effective potential as a function of the inhomogeneity strength. We have also demostarted the manifestation of the reentrant localization in the quench dynamics of the condensate. Our present analysis provides an alternate disorder parameter other than the external potential to attain the reentrant localization which may motivate the future experiments in the spinor BECs. 

In the present work, we have considered the equal intraspecies interaction between the components that restrict the same population among the components. It would be intriguing to extend the formalism developed in the present work to account for the effect of unequal intraspecies interactions on the localization and delocalization of the individual component(s). Another extension would include a completely random disordered potential, where the condensate could exhibit complex phases in the localized and delocalized state.{In another direction it would be interesting to extend the present formalism to analyze the localization-delocalization transition for the case of quasi-particles in the similar line as presented in the series of the works~\cite{Ray:2016, Lugan:2007,Lugan:2011}. }

\acknowledgments
R.R. acknowledges the postdoctoral fellowship supported by Zhejiang Normal University, China, under Grants No. YS304023964. T.M. acknowledges support from Science and Engineering Research Board (SERB), Govt. of India, through project No. MTR/2022/000382 and STR/2022/000023. The work of P.M. is supported by MoE RUSA 2.0 (Bharathidasan University - Physical Sciences). We gratefully acknowledge our super-computing facility Param-Ishan and Param Kamrupa (IITG), where all the simulation runs were performed.

%%%%%%%%%%%%%%%%%%%%%%%%%%%%%%%%%%%%%%%%%%%%
%%Appendix
%%%%%%%%%%%%%%%%%%%%%%%%%%%%%%%%%%%%%%%%%%%%
\appendix
\onecolumngrid
\counterwithin{figure}{section}
% \counterwithin{figure}{section}
\section{Details of the Variational approach Calculation}
\label{Appen:1}
\begin{figure}[!htp]
 \centering
 \includegraphics[width=0.5\linewidth]{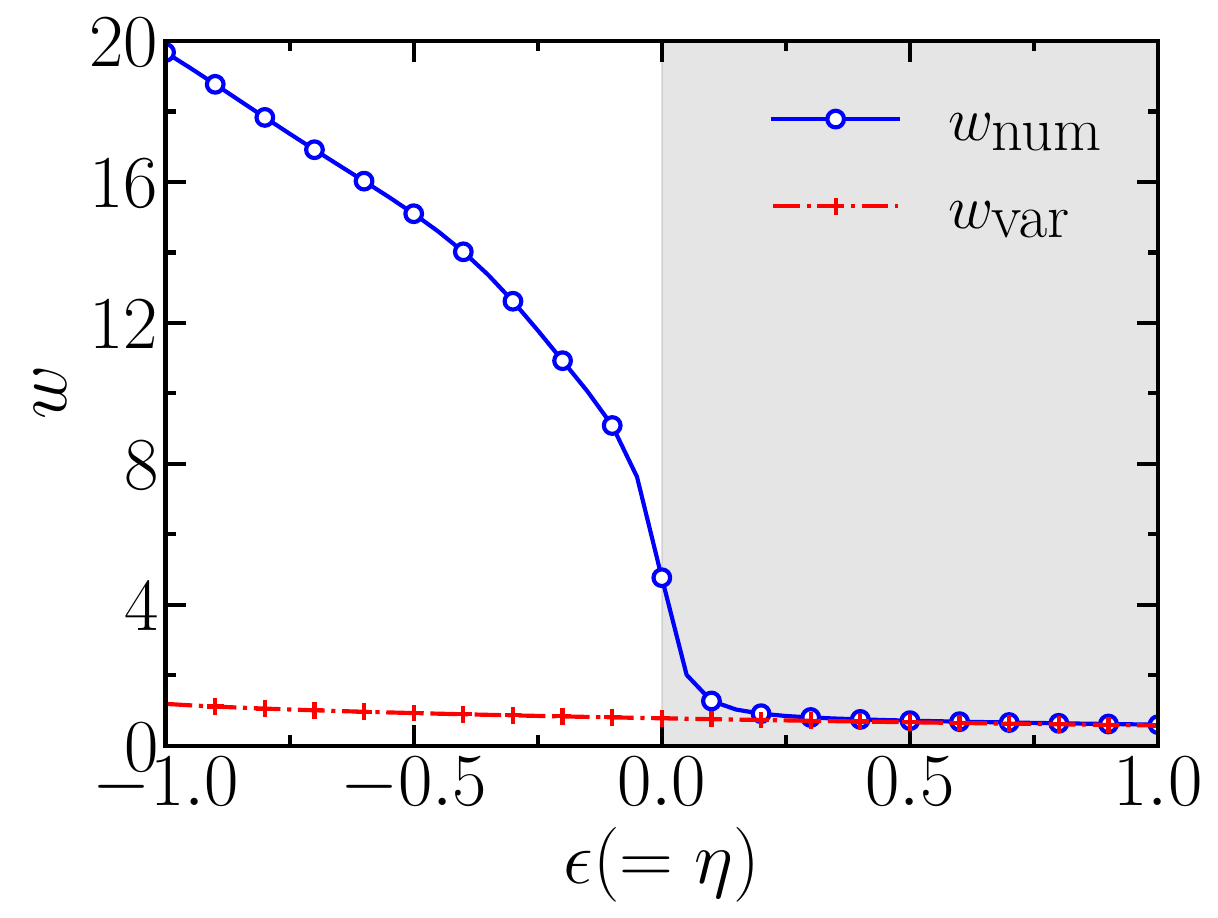}
 \caption{ Comparison of numerical $(w_{\rm num})$ and variational condensate width $(w_{\rm var})$ as a function of interaction inhomogeneity as $\epsilon(=\eta)$ at $k_L = 0.6,$ and $\Omega = 1$. $(w_{\rm num})$ (circle marked solid-blue line) obtained using the numerical integration of Eq.(\ref{eqn1(a)} - \ref{eqn1(b)}), and $(w_{\rm var})$ (red-dash dotted line) obtained using the solution of Eq.(\ref{eqn:A2b} - \ref{eqn:A2f}) with proper initial condition. In the localized region ($\epsilon(=\eta) > 0$), the $w_{\rm num}$ shows good agreement with the variational method while, in the delocalized region ($\epsilon(=\eta) < 0$), the variational method fails to accurately match the numerical condensate width.}
 \label{fig_appex:width-epsilon-eta}
 \end{figure}

%%%%%%%%%%%%%%%%%%%%%%%%%%%%%%%%%%%
In this appendix, we provide the detailed steps of time dependent variational approach. The Lagrangian for the zero phase difference between the inhomogeneous interaction and potential is given by
\begin{subequations}
 \begin{align}
% \begin{split}
 L = & \mathlarger{\sum_{j=1}^2} \Bigg[N_j (-1)^j\beta_j \dot{x}_{0_j} - N_j \dot{\phi_j} + N_j \beta_j k_L - \frac{N_j }{2}\left(\frac{1}{2w_j^2} + \beta_j^2\right) - \frac{N_j}{2^{3/2}\sqrt{\pi}w_j}\Bigg[\epsilon_0 - \epsilon \sum_{l=1}^{2} V_{l} \cos \left(k_{l} x_{0_j}\right) \exp \left(-\frac{k_{l}^{2} w_j^{2}}{8}\right)\Bigg] \notag \\ & + N_{j} \sum_{l=1}^{2} V_{l} \cos \left(k_{l} x_{0_j}\right) \exp \left(-\frac{k_{l}^{2} w_{j}^{2}}{4}\right) \Bigg] -\frac{N_{1} N_{2}}{\sqrt{\pi} \sqrt{w_1^{2}+w_2^{2}}} \exp \left(-\frac{\left(x_{01}-x_{02}\right)^{2}}{w_1^{2}+w_2^{2}}\right) \notag\\
 &  \times \Bigg[\eta_{0}-\eta \sum_{l=1}^{2} V_{l} \cos \left(\frac{k_{l}\left(w_1^{2} x_{0_{2}}+w_2^{2} x_{0_{1}}\right)}{w_1^{2}+w_2^{2}}\right)\exp \left(-\frac{k_{l}^{2} w_1^{2} w_2^{2}}{4\left(w_1^{2}+w_2^{2}\right)}\right)\Bigg] - 2\sqrt{N_1N_2}L_{\Omega} ,
% \end{split}
 \label{eqn:A1a}
 \end{align} 
%\end{subequations}
where
%\begin{subequations}
\begin{align}
L_{\Omega}=\Omega \sqrt{\frac{2 w_1 w_2}{\left(w_1^{2}+w_2^{2}\right)}} \cos \left[\frac{\left(\beta_{1} w_{1}^{2}-\beta_{2} w_{2}^{2}\right)\left(x_{0_1}-x_{0_{2}}\right)-\phi\left(w_1^{2}+w_2^{2}\right)}{w_1^{2}+w_2^{2}}\right] \exp \left( -\frac{(x_{0_1}- x_{0_2})^2 + (\beta_1 + \beta_2)^2w_1^2w_2^2}{2(w_1^2+w_2^2)}\right)
\label{eqn:A1b}
\end{align}
\end{subequations}
with $\phi = \phi_2 - \phi_1$, and $\rho = x_{0_1} - x_{0_2}$ 
%%%%%%%%%%%%%%%%%%%%%%%%%%%
The Euler-Lagrangian equation associated to variational parameters $\alpha_j$ is given by
\begin{subequations}
 \begin{align}
 \frac{\partial L}{\partial \alpha_j} - \frac{d}{dt}\frac{\partial L}{\partial \dot{\alpha_j}} = 0
 \label{eqn:A2a}
 \end{align}
%\end{subequations}
where $\alpha_j$ corresponds to the variational parameters $N_j$, $\beta_j$, $\phi_j$, $x_{0_j}$, and $w_j$.
%EOM with N_j
%\begin{subequations}
\vspace{1mm}
The Euler equation of motion corresponding to $\phi_j$ is gievn by
\begin{align}
	%\begin{split}
	\dot{\phi}_{j}= & -\frac{1}{4 w_j^{2}}+\frac{\beta_{j}^{2}}{2}- \frac{N_{j}}{\sqrt{2 \pi} w_j}\left[\epsilon_{0} - \epsilon \sum_{l=1}^{2} V_{l} \cos \left(k_{l} x_{0 j}\right) \exp \left(-\frac{k_{l}^{2} w_j^{2}}{8}\right)\right] 
	- \frac{N_{3-j}}{\sqrt{\pi\left(w_1^{2}+w_2^{2}\right)}} \exp \left(-\frac{\left(x_{0_{1}}-x_{0_{2}}\right)^{2}}{w_1^{2}+w_2^{2}}\right) \notag \\ 
	&  \left[\eta_{0} -\eta \sum_{l=1}^{2} V_{l} \cos \left(k_l\frac{w_1^{2} x_{0_2}+w_2^{2} x_{0_1}}{w_1^{2}+w_2^{2}}\right) \exp \left(-\frac{k_{l}^{2} w_1^{2} w_2^{2}}{4(w_{1}^{2}+w_2^{2})}\right)\right] - \sqrt{\frac{N_{3-j}}{N_j}}L_{\Omega} + \beta_j \sqrt{\frac{N_{3-j}}{N_j}} \frac{\partial L_{\Omega}}{\partial \beta_j} \notag \\
 &   + \sum_{l=1}^{2} V_{l} \cos \left(k_{l} x_{0_j}\right) \exp \left(-\frac{k_{l}^{2} w_{j}^{2}}{4}\right)  
	%\end{split}
	\label{eqn:A2b}
	\end{align}
In the similar line the equation of motion associated to $x_{0_j}$ is given by
%EOM with beta
\begin{align}
	\begin{split}
	\dot{x}_{0_j} = (-1)^{1-j}\left[- k_L + \beta_j + \sqrt{\frac{N_{3-j}}{N_j}}\hspace{2mm}\frac{\partial L_{\Omega}}{\partial \beta_j}\right]
	\end{split}
	\label{eqn:A2c}
	\end{align}
Further the equation of motion associated to $N_j$ can be written as:
%EOM with phi_j
\begin{align}
% \begin{split}
 \dot{N}_{j}=(-1)^{j} 2 \Omega \sqrt{N_{1} N_{2}} \sqrt{\frac{2 w_{1} w_{2}}{w_{1}^{2}+w_{2}^{2}}} \sin \left[\frac{\left(\beta_{1} w_{1}^{2}-\beta_{2} w_{2}^{2}\right)\left(x_{0_1}-x_{0_2}\right)}{\left(w_{1}^{2}+w_{2}^{2}\right)}-\phi\right]
 \exp \left[-\frac{\left(x_{0_1}-x_{0_2}\right)^{2}+\left(\beta_{1}+\beta_{2}\right)^{2} w_{1}^{2} w_{2}^{2}}{2\left(w_{1}^{2}+w_{2}^{2}\right)}\right]
% \end{split}
\label{eqn:A2d}
\end{align}
The Euler equation of motion corresponding to $\beta_j$ has the form as 
%EOM with x_0_j
	\begin{align}
	%\begin{split}
	\dot{\beta}_{j}= & \frac{N_{j} \epsilon}{(-1)^{3-j} 2  \sqrt{2\pi} w_{j}} \sum_{l=1}^{2}\left(V_{l} k_{l}\right) \sin \left(k_{l} x_{0_{j}}\right) \exp \left(-\frac{k_{l}^{2} w_j^{2}}{8}\right)
	+\frac{1}{(-1)^{3-j}} \sum_{l=1}^{2}\left(V_{l} k_{l}\right) \sin \left(k_{l} x_{0 j}\right) \exp \left(-\frac{k_{l}^{2} w_j^{2}}{4}\right) \notag \\
	& -\frac{2 \eta_{0} N_{3-j}\left(x_{0_{1}}-x_{0_{2}}\right)}{\sqrt{\pi}\left(w_{1}^{2}+w_{2}^{2}\right)^{3 / 2}}\exp \left(-\frac{\left(x_{0_{1}}-x_{0_{2}}\right)^{2}}{w_{1}^{2}+w_{2}^{2}}\right)
	-\frac{\eta N_{3-j}(-1)^{j-3}}{\sqrt{\pi}\left(w_{1}^{2}+w_{2}^{2}\right)^{3 / 2}} \exp \left(-\frac{\left(x_{0_{1}}-x_{0}\right)^{2}}{w_{1}^{2}+w_{2}^{2}}\right)\notag \\
	& \sum_{l=1}^{2}V_{l} \left[\left(-k_{l} w_{3-j}^{2}\right) \sin \left(k_l\frac{w_{1}^{2} x_{0_{2}}+w_{2}^{2} x_{0_{1}}}{w_{1}^{2}+w_{2}^{2}}\right)
	+(-1)^{j} 2\left(x_{0_{1}}-x_{0_{2}}\right)\cos \left(k_l\frac{w_{1}^{2} x_{0_{2}}+w_{2}^{2} x_{0_{1}}}{w_{1}^{2}+w_{2}^{2}}\right)\right] \notag \\
	& \times \exp \left(-\frac{k_{l}^{2} w_{1}^{2} w_{2}^{2}}{4\left(w_{1}^{2}+w_{2}^{2}\right)}\right)   \sqrt{\frac{N_{3-j}}{N_j}} (-1)^{j-3} \frac{\partial L_{\Omega}}{\partial x_{0_{j}}} - \frac{\dot{N}_{j}}{N_j} \beta_j 
	%\end{split}
	\label{eqn:A2e}
	\end{align}
%%%%%%%%%%%%%%%%
Finally, the Euler equation of motion with respect to $w_j$ is given by 
%EOM with omega_j
\begin{align}
 \begin{split}
 & \frac{N_{j}}{2 w_j^{3}} +\frac{N_{j}^{2}}{2 \sqrt{2\pi} w_{j}^{2}}\left[\epsilon_{0}-\frac{\epsilon}{4} \sum_{l=1}^{2} V_{l}\left(k_{l}^{2} w_j^{2}+4\right) \exp \left(-\frac{k_{l}^{2} w_j^{2}}{8}\right) \cos \left(k_{l} x_{0 j}\right)\right]-\frac{N_{j} w_j}{2} \sum_{l=1}^{2}\left(V_{l} k_{l}^{2}\right) \cos \left(k_{l} x_{0_{j}}\right) \exp \left(-\frac{k^2_{l} w^2_j}{4}\right)\\
 &  +\frac{N_{1} N_{2} \eta_{0} w_j}{\sqrt{\pi}\left(w_1^{2}+w_2^{2}\right)^{5 / 2}}\left(w_1^{2}+w_2^{2}-2\left(x_{0_{1}}-x_{0_{2}}\right)^{2}\right) \exp \left(-\frac{\left(x_{0_{1}}-x_{0_{2}}\right)^{2}}{w_1^{2}+w_2^{2}}\right) \\ 
 &  -\frac{\eta N_{1} N_{2} w_j}{2 \sqrt{\pi}\left(w_1^{2}+w_{2}^{2}\right)^{5 / 2}} \exp \left(-\frac{\left(x_{0_{1}}-x_{0_{2}}\right)^{2}}{w_1^{2}+w_2^{2}}\right)\sum_{l=1}^{2} V_{l}\bigg[\left(-4\left(x_{0_{1}}-x_{0_{2}}\right)^{2}+2 w_1^{2}+2 w_2^{2}+k_{l}^{2} \omega_{3-j}^{4}\right)
 \cos \left(k_l\frac{w_1^{2} x_{0_{2}}+w_2^{2} x_{0_{1}}}{w_1^{2}+w_2^{2}}\right) \\
 & +(-1)^{j} 4 k_{l}\left(x_{0_{1}}-x_{0_{2}}\right) \omega_{3-j}^{2} \sin \left(k_l\frac{w_{1}^{2} x_{0_{2}}+w_{2}^{2} x_{0_{1}}}{w_1^{2}+w_{2}^{2}}\right)\bigg]
 \exp \left(-\frac{k_{l}^{2} w_1^{2} w_2^{2}}{4\left(w_{1}^{2}+w_{2}^{2}\right)}\right)
 -\sqrt{N_{j}N_{3-j}} \frac{\partial L_{\Omega}}{\partial w_j}=0 
\end{split}
\label{eqn:A2f}
\end{align}
\end{subequations}
Note that the equation of motion corresponding to $w_j$ does not contain the time derivative of the condensate width $w_j$ as like the other equations (\ref{eqn:A2b})-(\ref{eqn:A2e}). Due to this reason, the variational approach does not seem to fit to analyze the dynamics of the width of the condensate. However, the condensate width obtained from the variational approach demonstrates good agreement in the localized region, as indicated by the shaded area in Fig.~\ref{fig_appex:width-epsilon-eta}. In contrast, the condensate in the delocalized region exhibits a spatially modulated stripe pattern (see Fig.~\ref{den-diff-kL}(a1)), which cannot be captured by the Gaussian variational approach.

%%%%%%%%%%%%%%%%%%%%%%%%%%%%%%%%%%%%%%%%%%%%%%%%%%%%%%%%%%%%%%%%%%%%%
%%%%%%%%%%%%%%%%%%%%%%%%%%%%%%%%%%%%%%%%%%%%%%%%%%%%%%%%%%%%%%%
%Note that, equation of motion corresponding to $w_j$ does not contain the time derivative of the condensate width $w_j$ as like the other equations (\ref{eqn:A2b}-\ref{eqn:A2e}). Due to this reason, variational approach does not seem deem fit to analyze the dynamics of the wodth of the condensate. 
%%%%%%%%%%%%%%%%%%%
%\appendix
%\subsection{Details of effective potential}
%\label{appen:2}
%%%%%%%%%%%%%%%%%%%%%%%%%%%%%%%%%%%%%%%%%%%%%%%%%%%%%%%%%%%%%%%
%\twocolumngrid

%%%%%%%%%%%%%%%%%%%%%%%%%%%%%%%%%%%%%%%%%%%%%%%%%%%%%%%%%%%%
Next, the detailed equation for the effective potential $V^{\rm eff}_j$ can be derived from the Eq.~(\ref{eqn:A2c}) and Eq.~(\ref{eqn:A2e}) and it will assume the form as 
\begin{subequations}
\begin{align}
V^{\mbox{eff}}_{j} = &-\frac{\epsilon}{2^{3 / 2} \sqrt{\pi} w_{j}} \sum_{l=1}^{2} V_{l}\cos \left(k_{l} x_{0_{j}}\right) \exp \left(-\frac{k_{l}^{2} w_j^{2}}{8}\right) %\notag \\ &
 - \sum_{l=1}^{2} V_{l}\cos \left(k_{l} x_{0_{j}}\right) \exp \left(-\frac{k_{l}^{2} w_j^{2}}{4}\right) %\notag \\ &
 + \frac{(-1)^{3-j}\eta_{0}}{\sqrt{\pi}\left(w_{1}^{2}+w_{2}^{2}\right)^{1 / 2}} \notag \\ 
& \times\exp \left(-\frac{\left(x_{0_{1}}-x_{0_{2}}\right)^{2}}{w_{1}^{2}+w_{2}^{2}}\right) 
 - \frac{\eta }{\sqrt{\pi} \sqrt{w_1^{2}+w_2^{2}}} 
 \exp \left(-\frac{\left(x_{01}-x_{02}\right)^{2}}{w_1^{2}+w_2^{2}}\right) 
 \sum_{l=1}^{2} V_{l} \cos \left(\frac{k_{l}\left(w_1^{2} x_{0_{2}}+w_2^{2} x_{0_{1}}\right)}{w_1^{2}+w_2^{2}}\right) \notag \\ & 
 \times \exp \left(-\frac{k_{l}^{2} w_1^{2} w_2^{2}}{4\left(w_1^{2}+w_2^{2}\right)}\right) %\notag \\ & 
 + L_{\Omega}
 \label{eqn:A3a}
\end{align}
\end{subequations}
The effective potential obtained in Eq.~(\ref{eqn:A3a}) is mainly composed of five different terms. The first term involving interactions $\epsilon$ contributes towards localization in case of repulsive interactions. The second term, containing the bichromatic optical lattice potential, is responsible for the localization of the condensate. The third term contains $\eta_0$, which determines the state of localization or delocalization depending upon the component and the sign of $\eta_0$. 
%Although, we consider $\eta_0 = 0$ throughout our study. The term containing $\eta$ behaves similarly in compare to the first term of this equation. 
The last term containing the Rabi-coupling [See Eq.(\ref{eqn:A3a})] contributes towards either localization or delocalization for in-phase or out-of-phase state of the condensate, respectively. For our calculation, we choose $\phi = \pi$ between two components in all cases which implies that the role of Rabi coupling towards localization. Further, the condensate width $w$, and chirp $\beta$ appearing in the effective potential can be estimated by solving Eqs.~(\ref{eqn8a}–\ref{eqn8c}). After substituting the $w_j$, and $\beta_j$, we compute the $V^{\rm eff}(x_0)$ for various spatial inhomogeneity and coupling parameters, which has been depicted in Fig.~\ref{fig:eff_potential}. 

%%%%%%%%%%%%%%%%%%%%%%%
\section{Details of the Lagrangian and effective potential for the phase shift $\pi/2$ between optical lattice potential and nonlinearity}
\label{appen:B}

\begin{figure}%[!htp]
\centering
\includegraphics[width=0.7\linewidth]{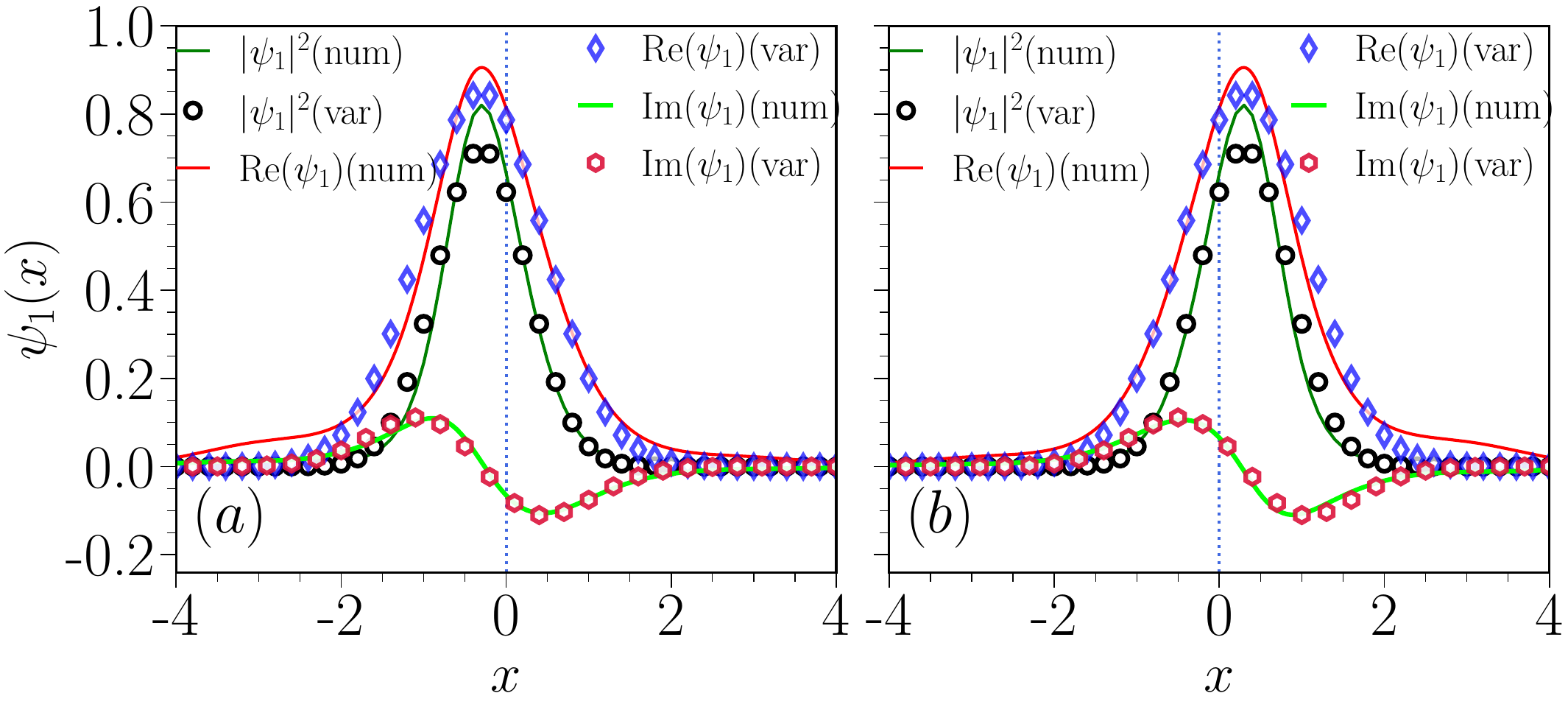}
\caption{
Comparison of the condensate density profile obtained numerically (solid line) and using the variational approach (open markers) for different inhomogeneity parameters when the phase difference between the nonlinear inhomogeneity and potential is $\pi/2$. (a) For $\epsilon =-\eta =-0.8$, and (b) For $\epsilon = - \eta = 0.8$. The other parameters are $k_L = 0.6, \theta_l = \pi/2k_l$, and $\Omega =1.0$. Incorporating the phase shift $\pi/2$ between the potential and inhomogeneity results in the condensate localizing at an off-center position at (a) $x = -0.3$ and (b) $x = 0.3$. The real (Re) and imaginary (Im) parts of the condensate wavefunction show good agreement between the wave function profile obtained from numerical simulation and that from using the variational approach.
%A comparison between the condensate density profile obtained numerically (solid line) and that using the variational approach (open markers) for different inhomogeneity prameters when the phase difference between the nonlinear inhomogeneity and potential is $\pi/2$. (a) For $\epsilon =-\eta =-0.8$, and (b) For $\epsilon = - \eta = 0.8$. The other parameters are $k_L = 0.6, \theta_l = \pi/2k_l$, and $\Omega =1.0$. Incorporating the phase shift $\pi/2$ between the potential and inhomogeneity results in the condensate localizing at an off-center position $x = -0.3$ in panel (a) and in (b) at $x = 0.3$. The real (Re) and imaginary (Im) parts of the condensate wavefunction show good agreement between the wave function profile obtained from numerical simulation and that from using the variational approach.
} 
 
\label{fig:density-compare-theta}
\end{figure}
In this appendix section, we provide the Lagrangian and the corresponding equation of the effective potential for the case of $\theta = \pi / 2$ between optical lattice and interaction inhomogeneity. Here, $j = 1, 2$ represents the spin-up and spin-down components, respectively. 
%%%%%%%%%%%%%%%%%%%%%%%%%%%%%%%%%%%%%%%%%%%%%%%%%%%%%%%%%%%%%
\begin{figure}[!htp]
\centering
\includegraphics[width=0.6\linewidth]{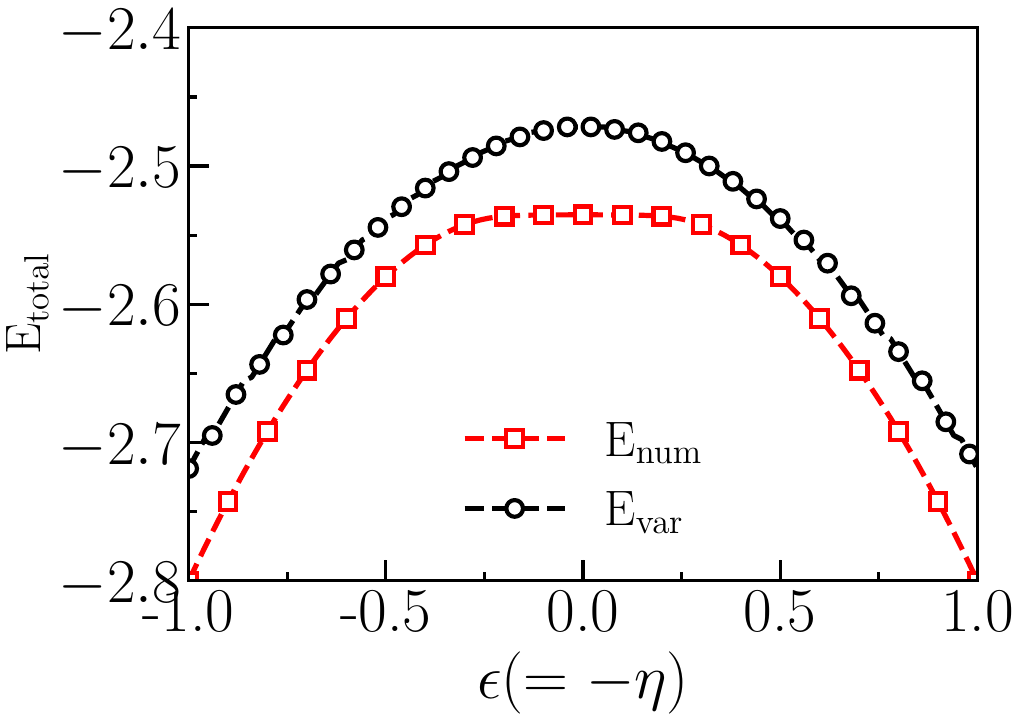}
\caption{Variation of numerical ($E_{\rm num}$) and variational ($E_{\rm var}$) total energy with inhomogeneity $\epsilon(=-\eta)$ with $k_L = 0.6$, and $\Omega = 1.0$. Tuning the inhomogeneity leads to a monotonic decrease in the total energy in the localized regions, that is $-1 \lesssim \epsilon(=-\eta) \lesssim -0.3$ or $0.3 \lesssim \epsilon(=-\eta) \lesssim 1.0$ whereas, in the delocalized region $-0.3 \lesssim \vert \epsilon(=-\eta)\vert \lesssim 0.3$, the total energy remains almost constant at $E_{\rm num} \sim -2.56$. In both cases, energy follow a symmetric nature in either side of delocalization regions. }
\label{fig:Energy-piby2}
\end{figure}
%%%%%%%%%%%%%%%%%%%%%%%%%%%%%%%%%%%%%%%%%
%
%\onecolumngrid

\begin{subequations}
\begin{align}
 \begin{split}
 L = & \mathlarger{\sum_{j=1}^2} \Bigg[N_j (-1)^j\beta_j \dot{x}_{0_j} - N_j \dot{\phi_j} + N_j \beta_j k_L - \frac{N_j }{2}\left(\frac{1}{2w_j^2} + \beta_j^2\right) - \frac{N_j}{2^{3/2}\sqrt{\pi}w_j}\Bigg[\epsilon_0 - \epsilon \sum_{l=1}^{2} V_{l} \sin \left(k_{l} x_{0_j}\right) \exp \left(-\frac{k_{l}^{2} w_j^{2}}{8}\right)\Bigg] \\ & \left. + N_{j} \sum_{l=1}^{2} V_{l} \cos \left(k_{l} x_{0_j}\right) \exp \left(-\frac{k_{l}^{2} w_{j}^{2}}{4}\right)\Bigg] -\frac{N_{1} N_{2}}{\sqrt{\pi} \sqrt{w_1^{2}+w_2^{2}}} \exp \left(-\frac{\left(x_{01}-x_{02}\right)^{2}}{w_1^{2}+w_2^{2}}\right) \right. \\
 & \left. \times \Bigg[\eta_{0}+\eta \sum_{l=1}^{2} V_{l} \sin \left(\frac{k_{l}\left(w_1^{2} x_{0_{2}}+w_2^{2} x_{0_{1}}\right)}{w_1^{2}+w_2^{2}}\right)\exp \left(-\frac{k_{l}^{2} w_1^{2} w_2^{2}}{4\left(w_1^{2}+w_2^{2}\right)}\right)\Bigg] - 2\sqrt{N_1N_2}L_{\Omega} \right.
 \end{split}
 \label{eqn:B1a}
 \end{align}
In the above equation $L_{\Omega}$ has the same form as in Eq.~(\ref{eqn:A1b}). The effective potential $V_{\rm eff}$ assumes the form as, 
\begin{align}
 V^{\mbox{eff}}_j =  &-\frac{\epsilon}{2^{3 / 2} \sqrt{\pi} w_{j}} \sum_{l=1}^{2} V_{l}\sin \left(k_{l} x_{0_{j}}\right) \exp \left(-\frac{k_{l}^{2} w_j^{2}}{8}\right) %\notag \\ &
 - \sum_{l=1}^{2} V_{l}\cos \left(k_{l} x_{0_{j}}\right) \exp \left(-\frac{k_{l}^{2} w_j^{2}}{4}\right) %\notag \\ &
 - \frac{\eta_{0}}{\sqrt{\pi}\left(w_{1}^{2}+w_{2}^{2}\right)^{1 / 2}} \notag \\ &
 \times \exp \left(-\frac{\left(x_{0_{1}}-x_{0_{2}}\right)^{2}}{w_{1}^{2}+w_{2}^{2}}\right) %\notag \\ &
 + \frac{\eta }{\sqrt{\pi} \sqrt{w_1^{2}+w_2^{2}}} 
 \exp \left(-\frac{\left(x_{01}-x_{02}\right)^{2}}{w_1^{2}+w_2^{2}}\right) %\notag \\ &
 \sum_{l=1}^{2} V_{l} \sin \left(\frac{k_{l}\left(w_1^{2} x_{0_{2}}+w_2^{2} x_{0_{1}}\right)}{w_1^{2}+w_2^{2}}\right) \notag \\ &
 \times \exp \left(-\frac{k_{l}^{2} w_1^{2} w_2^{2}}{4\left(w_1^{2}+w_2^{2}\right)}\right)  + L_{\Omega}
 \label{eqn:B1b}
\end{align}
\end{subequations}
%%%%%%%%%%%%%%

In Fig. \ref{fig:density-compare-theta}(a) and Fig. \ref{fig:density-compare-theta}(b), we show a comparison between the density profiles obtained from the numerical simulation (solid lines) and that from the variational approach (marked with open markers) for $\epsilon=-0.8$, $\eta = 0.8$ and $\epsilon=0.8$, $\eta = -0.8$, respectively. For both the combination of the strength of the inhomogeneity, the condensate gets localized at off-center. For instance, in Fig.~\ref{fig:density-compare-theta}(a), the condensate gets localized at $x \approx -0.3$, while, in Fig.~\ref{fig:density-compare-theta}(b), the localization happens at $x \approx 0.3$. To unravel the reason for this particular nature of localization of the condensate at the off-center position in space, we analyze the nature of effective potential as depicted in Fig.~\ref{fig:eff_potential_theta}(a). We find that increasing the strength of interaction results in shifting the minima of the effective potential $V^{\rm eff}$ either left or right, which depends solely on the nature of the interaction inhomogeneity. The left shift happens for the attractive ($\epsilon=-\eta = - 1.0$) while the right shift happens for the repulsive intraspecies ($\epsilon=-\eta = 1.0$) interactions. 
%In Fig. \ref{fig:density-compare-theta}(a) and Fig. \ref{fig:density-compare-theta}(b) we show a comparison between the density profiles obtained from the numerical simulation (solid lines) and that from the variational approach (marked with open markers) for $\epsilon=-0.8$, $\eta = 0.8$ and $\epsilon=0.8$, $\eta = -0.8$, respectively. For both the combination of the strength of the inhomogeneity the condensate gets localized at off-center. For instance, in Fig.~\ref{fig:density-compare-theta}(a), the condensate gets localized at $x \approx -0.3$, while, in Fig.~\ref{fig:density-compare-theta}(b), the localization happens at $x \approx 0.3$. To unravel the reason this particular nature of localization of the condensate at the off-center position in space, we analyze the nature of effective potential as depicted in Fig.~\ref{fig:eff_potential_theta}(a). We find that upon increasing value the strength of interaction results shifting the minima of the effective potential $V^{\rm eff}$ either left or right which depends solely on the nature of the interaction inhomgeneity. Left shift happens for the attractive ($\epsilon=-\eta = - 1.0$) while right shift happens for the repulsive intraspecies ($\epsilon=-\eta = 1.0$) interactions. 
%The shifting of the lowest minima clearly demonstrates the localization of the condensate at the off-center position. 

To get more insight into the probable cause for the localization and delocalization transition, in Fig.~\ref{fig:Energy-piby2}, we show a variation of the total energy as a function of the interaction inhomogeneity obtained from the numerical simulation and using the variational approach. The energy shows a monotonically decreasing trend with the interaction inhomogeneity strength for $0.3 \lesssim \lvert \epsilon(=-\eta)\rvert \lesssim 1$. However, for $0 \lesssim \lvert\epsilon(=-\eta)\rvert \lesssim 0.3$ the energy assumes the constant value around $E_{\rm num} \approx -2.56$. Variational energy $E_{\rm var}$ qualitatively follows a similar trend as that of $E_{\rm num}$ in both localized and delocalized regions. 

\begin{figure}[H]
    \centering
    \includegraphics[width=0.8\linewidth]{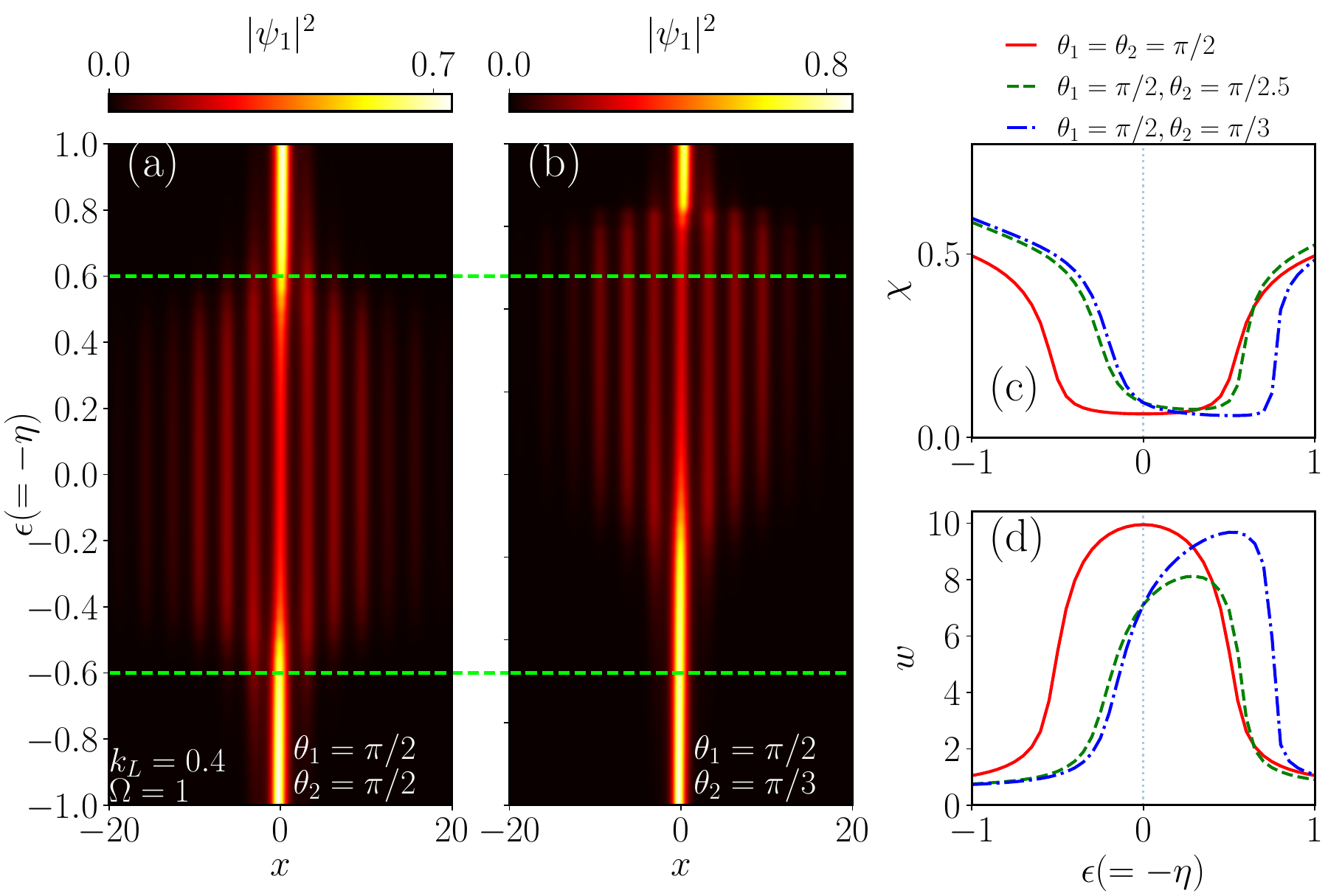}
    \caption{Pseudocolor representation of the spatial variation of the ground state density of the spin-up component ($|\psi_1|^2$) with the inhomogeneity parameter $\epsilon(=-\eta)$ for (a) $\theta_1 = \theta_2 = \pi/2$, and (b) $\theta_1 = \pi/2$, $\theta_2 = \pi/3$. Upon choosing $\theta_1 = \theta_2$ results in the symmetric localized and delocalized regions about $\epsilon = -\epsilon$ while $\theta_1 \neq \theta_2$ leads to break that symmetry about $\epsilon = -\epsilon$ as shown in (b). Variation of (c) form-factor $\chi$, and (d) width $w$ of the condensate as a function of interaction inhomogeneity $\epsilon=(-\eta)$. Upon inclusion of unequal $\theta_1$, and $\theta_2$ in interaction inhomogeneity leads to break the symmetry about $\epsilon = -\epsilon$. The coupling parameters are considered at $k_L = 0.4$, $\Omega = 1.0$. The green dashed line is drawn to compare the symmetric delocalized region in Figure (b) with (a). }
    \label{fig:1}
\end{figure}

Next, in Fig.~\ref{fig:1}(a,b), we show the pseudocolor density profile representation as a function of $(\epsilon=-\eta)$ for the cases $\theta_1=\theta_2=\pi/2$ and $\theta_1=\pi/2, \theta_2=\pi/3$, respectively. At one hand for $\theta_1=\theta_2=\pi/2$ the localization-delocalization transition takes place for $\epsilon\rightarrow-\epsilon$, $\eta \rightarrow -\eta$, but for $\theta_1=\pi/2$, $\theta_2=\pi/3$ the transition is not symmetrical about  ($\epsilon\rightarrow-\epsilon, \eta \rightarrow -\eta$) [see Fig.~\ref{fig:1}(b)]. To make our claim more concrete in Figs.~\ref{fig:1}(c,d), we plot the form-factor $\chi$ and the width $w$ of the condensate, for various combination of the phase factor in the interaction inhomogeneity, such as (a) $\theta_1=\theta_2=\pi/2$, (b) $\theta_1=\pi/2$, $\theta_2=\pi/2.5$, as (c) $\theta_1=\pi/2$, $\theta_2=\pi/3$. It is evident that upon choosing $\theta_1\ne \theta_2$, the symmetry in the appearance of reentrant localization with respect to the change in the sign of the interaction of the inhomogeneity parameter ceases to exist. Therefore, the symmetrical behavior of localization-delocalization transition can be broken by choosing different phase shift between the interaction inhomogeneity and the quasiperiodic potential.

\twocolumngrid

\bibliography{references}

\end{document}